\documentclass[aps,nofootinbib,groupedaddress,twocolumn,notitlepage,superscriptaddress,10pt]{revtex4-1}

\usepackage{float}
\usepackage{color}
\usepackage{bbm}
\usepackage{tikz}

\newcommand{\nb}[1]{\color{blue}}

\newcommand{\hl}[1]{\color{magenta}}

\usepackage[
pagebackref=false,
colorlinks=true,
linkcolor=blue,
urlcolor=blue,
filecolor=black,
citecolor=red,
pdfstartview=FitV,
pdftitle={},
pdfauthor={},
pdfsubject={},
pdfkeywords={},
pdfpagemode=None,
bookmarksopen=true
]{hyperref}

\usepackage[normalem]{ulem}
\usepackage{amsmath, amssymb}
\usepackage{enumerate}
\usepackage{amsfonts}
\usepackage{epsfig}
\usepackage{mathbbol}
\usepackage{mathrsfs}
\usepackage{braket}

\newcommand{\be}{\begin{equation}}
\newcommand{\ee}{\end{equation}}
\newcommand{\bea}{\begin{eqnarray}}
\newcommand{\eea}{\end{eqnarray}}
\newcommand{\bega}{\begin{gather}}
\newcommand{\eega}{\end{gather}}

\newcommand{\bi}{\begin{itemize}}
\newcommand{\ei}{\end{itemize}}
\newcommand{\ben}{\begin{enumerate}}
\newcommand{\een}{\end{enumerate}}
\newcommand{\bca}{\begin{cases}}
\newcommand{\eca}{\end{cases}}
\newcommand{\bln}{\begin{align}}
\newcommand{\eln}{\end{align}}
\newcommand{\bst}{\begin{split}}
\newcommand{\est}{\end{split}}
\def\ie{\begin{equation}\begin{aligned}}
\def\fe{\end{aligned}\end{equation}}
\newcommand{\bma}{\le(\begin{matrix}}
\newcommand{\ema}{\end{matrix}\ri)}

\def\le{\left}
\def\ri{\right}

\definecolor{darkred}{rgb}{0.8,0.1,0.2}

\begin{document}

\title{\textbf{Gauging anomalous unitary operators}}

\author{Yuhan Liu}
\affiliation{Kadanoff Center for Theoretical Physics, University of Chicago, Chicago, Illinois 60637, USA}
\affiliation{James Franck Institute, University of Chicago, Chicago, Illinois 60637, USA}
 
\author{Hassan Shapourian%
}
\affiliation{Microsoft Station Q, Santa Barbara, California 93106, USA}

\author{Paolo Glorioso}
\affiliation{Department of Physics, Stanford University, Stanford, California 94305, USA}

\author{Shinsei Ryu}
\affiliation{ Department of Physics, Princeton University, Princeton, New Jersey 08540, USA}

\begin{abstract}
  Boundary theories of static bulk
 topological phases of matter
  are obstructed in the sense that they cannot be realized
  on their own as isolated systems. 
  The obstruction can be quantified/characterized
  by quantum anomalies, in particular when there is a global symmetry.
  Similarly, topological Floquet evolutions
  can realize obstructed unitary operators at their boundaries.
  In this paper, we discuss the characterization of
  such obstructions by using quantum anomalies.
  As a particular example, we discuss time-reversal symmetric
 boundary unitary operators in one and two spatial dimensions,
  where the anomaly emerges as we gauge the so-called Kubo-Martin-Schwinger (KMS) symmetry.
  We also discuss mixed anomalies between particle number conserving
  $U(1)$ symmetry and discrete symmetries, such as $C$ and ${\it CP}$,
  for unitary operators in odd spatial dimensions
  that can be realized
  at the boundaries of topological Floquet systems 
  in even spatial dimensions.
\end{abstract}

\maketitle
{\hypersetup{linkcolor=black} \tableofcontents}

\section{Introduction}

As the ground states of static, gapped Hamiltonians, unitary time-evolution
operators of quantum many-body systems can be topologically distinct from each
other or may exhibit topological properties.
For example, time-evolution operators of periodically-driven systems (Floquet systems) can give rise to
Floquet Hamiltonians that 
are topological much the same way as static topological systems, 
and
also
novel out-of-equilibrium phases of matter
that do not have static counterparts
\cite{2009PhRvB..79h1406O,
  2010PhRvB..82w5114K,
  2011NatPh...7..490L,  
  2011PhRvL.106v0402J,
  2018arXiv180403212O,
  2020NatRP...2..229R,
  2019arXiv190501317H}.
Floquet topological systems have been experimentally realized
in synthetic systems, such as ultracold atoms, photonic, and phononic systems
-- see for example
\cite{peng2016experimental,reichl2014floquet,cheng2019observation,2017RvMP...89a1004E}.

Similar to static topological phases,
some Floquet unitaries 
are topological even in the absence of any symmetry,
while others 
are topological in the presence of some symmetry,
i.e., their topological properties (topological distinction)
are protected by a symmetry.
The examples of the former 
include those that
support unidirectional quantum information flow
at their boundaries,
and are characterized by the chiral unitary index
(GNVW index)
\cite{rudner2013anomalous,po2016chiral,2017arXiv170307360F,2017PhRvL.118k5301H}.
On the other hand, 
bosonic Floquet systems
in $d$ spatial dimensions
with a symmetry group $G$
are classified by group cohomology
$H^{d+1}(\tilde{G},U(1))$
where $\tilde{G}=G\times \mathbb{Z}$
or $G\rtimes \mathbb{Z}$
\cite{2016PhRvB..93x5145V,
  2016PhRvB..93t1103E,
  2016PhRvX...6d1001P,
  2017PhRvB..95s5128R}.
For non-interacting fermion systems,
non-trivial topological Floquet unitaries
in the ten Altland-Zirnbauer symmetry classes
have been classified 
\cite{2017PhRvB..96o5118R,2017PhRvB..96s5303Y}.

In static topological phases,
it is known that
a bulk-boundary correspondence holds. 
The boundary theory of a bulk topological phase is anomalous,
in that it cannot be realized on its own as a local consistent theory.
For example, on the boundary of a bulk
symmetry-protected 
topological (SPT) phase 
protected by a global on-site symmetry,
the symmetry cannot act purely locally (i.e., the symmetry action is neither on-site
nor splittable); the boundary theory suffers from a 't Hooft anomaly.
In general, quantum anomalies at the boundary go hand in hand with non-trivial bulk topology,
and can be used as a diagnosis of the corresponding bulk.
Such anomalies can often be detected by gauging, i.e., by subjecting the boundary theory
to a background gauge field associated with the symmetry group
\cite{2008PhRvB..78s5424Q,
  2013PhRvB..87o5114C,
  2014arXiv1403.1467K,
  2014arXiv1404.6659K,
  2015JHEP...12..052K,
  2012PhRvB..86k5109L,
  2012PhRvB..85x5132R,
  PhysRevB.88.075125,
  2017PhRvB..95p5101S,
  2016RvMP...88c5005C}.
One natural question is whether a similar formalism is applicable to Floquet topological phases. 

In this paper, we discuss the anomalous (or topological) properties of
unitary time-evolution operators that 
may appear on the boundary of topological Floquet unitary operators.
In one-spatial dimension for bosonic systems,
these unitaries (locality-preserving quantum cellular automata)
can be expressed in terms of matrix-product unitaries
\cite{
  2017JSMTE..08.3105C,
  2018PhRvB..98x5122S,
  2018arXiv181201625H,
  2018arXiv181209183G,
  2019arXiv190210285F,
  2020arXiv200715371P,
  2020arXiv200711905P}.
We consider these unitaries in the presence of
a global symmetry, including 
discrete symmetries, such as time-reversal,
parity (reflection), charge-conjugation,
and combinations thereof.
In particular, we will develop gauging procedures,
i.e., to introduce background gauge fields, 
to detect anomalous properties of these unitaries.
As we will show,
the boundary unitaries of topological Floquet systems
suffer from quantum anomalies of discrete symmetries,
similar to the boundary states appearing in static topological phases.
The gauging procedure leads to explicit forms (formulas)
of (many-body) topological invariants that can be used for
arbitrary unitary operators with symmetries.

One convenient way to formulate  
our gauging procedure is to use the operator-state map
(reviewed in Sec.\ \ref{The operator-state map and KMS condition}),
and regard unitary operators as
short-range entangled states in the doubled Hilbert space,
which may be viewed as unique ground states of some gapped
Hamiltonians.
We can then use tools from the physics of
symmetry-protected topological phases to study the mapped states;
we can follow the gauging procedure for static topological phases of matter.
Section \ref{Gauging symmetries} is devoted to developing this idea.
In particular, we will also establish
the connection between the gauging procedure
with temporal background gauge fields and 
the approach in Ref.\ \cite{2016PhRvB..93x5145V}
that deals with anomalous operator algebras appearing
on boundaries of 1d topological Floquet systems. 
We will then generalize
to incorporate spatial components of background gauge fields.

Also in Sec.\ \ref{Gauging symmetries},
we will discuss how we can gauge
time-reversal symmetry.
Specifically,
as we will review in Sec.\ \ref{The operator-state map and KMS condition}, 
in the Schwinger-Keldysh formalism
or in thermofield dynamics,
time-reversal symmetry can be implemented as 
a unitary on-site $\mathbb{Z}_2$ symmetry 
-- this symmetry is the so-called KMS (Kubo-Martin-Schwinger) symmetry
\cite{2015PhRvB..92m4307S,2018arXiv180509331G, 2020arXiv200710448A}.
This symmetry can be gauged in much the same way as
unitary on-site symmetries in static topological phases of matter,
in order to diagnose
topological/anomalous properties of unitary operators.

We will apply the gauging procedure to diagnose anomalous (topological) properties of
matrix product unitaries (Sec.\ \ref{Matrix product unitaries}),
and boundary unitaries of Floquet Majorana fermion systems
(Sec.\ \ref{Majorana fermion models}).
For the case of 1d Majorana unitaries
(realized at the boundaries of 2d Floquet topological unitaries),
the model of our interest can be constructed by combining
two copies of the Majorana fermion model
with opposite chiralities discussed in \cite{2017arXiv170307360F}.
We will also discuss
2d time-reversal symmetric Majorana unitaries
that can be realized on the boundary of
3d bulk topological Floquet unitaries.
Gauging the KMS symmetry reveals 
the $\mathbb{Z}_8$ classification of these unitaries.

In Sec.\ \ref{Anomalous unitary operators with $U(1)$ and discrete symmetries},
we will consider the boundary unitatires of
Floquet topological systems of charged fermions. 
Namely, there is a $U(1)$ charge $Q$ which commutes with
these unitaries, $e^{i \theta Q}U e^{-i \theta Q}=U$ ($\theta \in [0,2\pi]$).
The examples include 
the 1d boundary unitary
of 2d Floquet topological Anderson insulators
\cite{rudner2013anomalous, 2016PhRvX...6b1013T,
  2017PhRvL.119r6801N,
  2019arXiv190712228N,
  2020arXiv201002253Z,
  2019arXiv190803217G}.
As shown in \cite{2019arXiv190803217G} the 1d boundary unitaries suffer from
a mixed anomaly between $U(1)$ and particle-hole symmetry.
In this paper, we extend this analysis
to higher-dimensional examples,
and show that
the anomalies are characterized by the Chern-Simons forms.
This is analogous to the dimensional hierarchy 
of topological response theories of topological insulators
discussed in Refs.\ \cite{2008PhRvB..78s5424Q,2010NJPh...12f5010R}.
We also construct many-body topological invariants
that can extract the Chern-Simons forms.

\section{The operator-state map and KMS condition}
\label{The operator-state map and KMS condition}

In this section, we will go through the ingredients of the operator-state map and the KMS symmetry
that are necessary for our analysis of unitary operators
\cite{Umezawa:1993yq,Ojima:1981ma}.
%

\subsection{The operator-state map}

\paragraph{The reference state}

We begin by reviewing some essential points of the operator-state map,
which maps operators acting on a Hilbert space $\mathcal{H}$
to the corresponding states in the doubled Hilbert space $\mathcal{H}\otimes
\mathcal{H}$ -- see below.
In broader contexts, one can apply the channel-state map (the Choi-Jamio\l kowski isomorphism)
to arbitrary quantum channels (trace-preserving completely positive map),
and associate them with quantum states (density matrices)
in the doubled Hilbert space.

We start from the identity operator $I = \sum_i |i\rangle \langle i|$,
and normalize it as
$\Omega = \sum_i |i\rangle \langle i|/ \sqrt{\mathcal{N}}$
so that $\mathrm{Tr}\, [\Omega^{\dag}\Omega] =1$.
Here, $\mathcal{N}=\mathrm{dim}\, \mathcal{H}=\mathrm{Tr}\, I$ is the dimension of the Hilbert space. 
By ``flipping'' the bras in $\Omega$, we define a ``reference state'',
a maximally entangled state in the doubled Hilbert space $\mathcal{H}\otimes \mathcal{H}^*$:
\begin{align}
  \label{ref state}
  |\Omega \rangle\!\rangle
\equiv
 ({1}/{\sqrt{\mathcal{N}}})
  \sum_i
  |i \rangle \otimes
  |i \rangle^*.
\end{align}
Here, $|i\rangle^* = K|i\rangle$
transforms as a conjugate representation
where $K$ is complex conjugation.
Under a unitary transformation $V$ acting on $\mathcal{H}$,
$|i\rangle$ and $|i\rangle^*$
transform complementarily as
\begin{align}
  \label{unitary trs}
  &
  |i \rangle \to
    V|i\rangle
    =
  \big(
  \sum_j |j\rangle \langle j|
  \big)
  V|i \rangle
  =
  \sum_j |j\rangle V_{ji},
  \nonumber \\
  &
    |i \rangle^*
    \to
    K VK^{-1} K|i\rangle =
    K V |i\rangle
    =
    \sum_j |j\rangle^* V_{ji}^*,
\end{align}
i.e., 
$|i\rangle$ and $|i\rangle^*$ transform as the fundamental and anti-fundamental representations of $U(\mathcal{N})$, respectively.
We refer to these two Hilbert spaces as ``out'' and ``in'' Hilbert spaces.
Like the operator $\Omega$,
which is invariant under a unitary transformation on $\mathcal{H}$,
$ \Omega \to V \Omega V^{\dag}
= \sqrt{\mathcal{N}}^{-1}\sum_i V|i\rangle \langle i|V^{\dag}
=
\Omega$,
the reference state $|\Omega\rangle\!\rangle$ 
enjoys the invariance under \eqref{unitary trs}.
The point here is that we consider the product
of two representations that are conjugate to each other.
The resulting product representation 
always includes a singlet representation.
(While we use complex conjugation $K$ here to pair up two Hilbert spaces, 
in later examples, we will consider a physical antiunitary symmetry operation,
such as time-reversal or time-reversal combined with
a unitary symmetry, such as $CT$, to define conjugate kets.)

\paragraph{The operator-state map}

We can now introduce the operator-state map
using the reference state $|\Omega \rangle\!\rangle$.
Let us consider a unitary operator $U$ acting on $\mathcal{H}$.
We introduce a state $|U\rangle\!\rangle$ corresponding to $U$ as
\begin{align}
  |U\rangle\!\rangle
  =
  (U \otimes I)|\Omega\rangle\!\rangle.
\end{align}
It is customary 
to 
represent 
the state-operator map
diagramatically as:
$$
\begin{tikzpicture}[thick, scale=.4]
  \begin{scope}
    \node at (0,3.5) {out};
    \node at (0,-0.5) {in};
  \draw (0,1) -- (0,0);
  \draw (0,2) -- (0,3);
  \node at (0,1.5) {$U$};
    \node at (3.8,3.5) {out};
    \node at (6,3.5) {in};
  \draw (-0.5,1) rectangle (0.5,2);

  \draw (3,1.5) node[left] {$\longrightarrow$};

  \draw (4,3) -- (4,2);
  \draw (4,1) -- (4,0) -- (5,-0.5) -- (6,0) -- (6,3);
  \node at (4, 1.5) {$U$};
  \draw (3.5,1) rectangle (4.5,2);
  \node at (5, -1) {$|\Omega\rangle\!\rangle$};
  \end{scope}
\end{tikzpicture}
$$
Here, the reference state $|\Omega \rangle\!\rangle$ appears as a ``cup".

Note that the overlap of two states corresponding to
unitaries $U$ and $U'$ can be written as
\begin{align}
  \label{overlap = SK}
\langle \!\langle U |U'\rangle\!\rangle
  =
  (1/{\cal N})
  \mathrm{Tr} \left[ U^{\dag} U^{\prime} \right]
  =
  \mathrm{Tr} \left[ U^{\prime}\rho_0 U^{\dag} \right],
\end{align}
where
the trace is taken over the original (single) Hilbert space $\mathcal{H}$, and
$\rho_{0}= I/{{\cal N}}$ is the infinite temperature state.
This overlap can be represented as a Schwinger-Keldysh path-integral
with the infinite temperature thermal state as the initial state.

\paragraph{The shift property}

Using the invariance of the reference state under $U$,
$|\Omega\rangle\!\rangle=
(U^{\dag}\otimes K U^{\dag}K^{-1})|\Omega \rangle\!\rangle
$,
the state $|U\rangle\!\rangle$ can also be written as
\begin{align}
  |U\rangle\!\rangle
  &=
  (U\otimes I)(U^{\dag}\otimes KU^{\dag}K^{-1}) |\Omega\rangle\!\rangle
    \nonumber \\
  &=
  (I\otimes KU^{\dag}K^{-1}) |\Omega\rangle\!\rangle
\end{align}
I.e., one can ``shift'' $U$ from the left (out) to right (in),
by conjugating with $K$.
This reflects 
the fact that
acting an arbitrary operator $O$ on $\Omega$ from the right and left give the identical result, 
$
 O= O\cdot \Omega = \Omega \cdot O
 =
 \sqrt{{\cal N}}^{-1}
 \sum_i O|i\rangle \langle i|
 =
 \sqrt{{\cal N}}^{-1}
 \sum_i |i\rangle \langle i| O.
$
The shift property of $|\Omega \rangle\!\rangle$
can be represented pictorially as
$$
\begin{tikzpicture}[thick, scale=.4]
  \begin{scope}
    \draw (0,1) -- (0,0) -- (1,-0.5) -- (2,0) -- (2,3); \draw (0,2) -- (0,3);
    \node at (0,1.5) {$U$};
    \draw (-0.5,1) rectangle (0.5,2);

    \draw (3.5,1.5) node[left] {$=$};

    \draw (4,3) -- (4,0) -- (5,-0.5) -- (6,0) -- (6,1); \draw (6,2) -- (6,3);
    \node at (6.4, 1.5) {$K U^{\dag}K^{-1}$}; \draw (4.5,1) rectangle (8.3,2);
  \end{scope}
\end{tikzpicture}
$$

\paragraph{The modular conjugation}

In the language of Tomita-Takesaki theory, the state operator map naturally comes with
an antiunitary operator acting on the doubled Hilbert space,
called the modular conjugation operator, which we denote by $J$.
Intuitively, $J$ can be understood as an
operation that exchanges the system of our interest and ``heat bath'';
in our case, it is an operation that exchanges the in and out Hilbert spaces.
As we will see, the KMS condition, within the framework of the thermofield dynamics,
can be stated by using $J$.
For the setting we are working with, $J$ can be introduced as
\begin{align}
  J \big(|i\rangle |j\rangle^*\big) =
  |j\rangle |i\rangle^*,
\end{align}
i.e.,
$J = K\cdot {\it SWAP}$,
where $K$ is complex conjugation acting on
$\mathcal{H}_{{\it out}}\otimes \mathcal{H}_{{\it in}}$,
and
${\it SWAP}$ exchanges the in and out Hilbert spaces. 
Note that the reference state $|\Omega\rangle\!\rangle$
is invariant under $J$.
The modular conjugation $J$ acts
on $|U\rangle\!\rangle$ as
\begin{align}
  \label{J symmetry}
  J |U\rangle\!\rangle
  &=
  J (U\otimes I) J\cdot J
    |\Omega \rangle\!\rangle
    \nonumber\\
    &=
    (I \otimes KUK^{-1})
    |\Omega\rangle\!\rangle
    \nonumber \\
    &=
    (U^{\dag}\otimes I)|\Omega \rangle\!\rangle
      =
      |U^{\dag}\rangle\!\rangle.
\end{align}
Here, we used the shift property of $|\Omega\rangle\!\rangle$. Diagramatically,
$$
\begin{tikzpicture}[thick, scale=.4]
  \begin{scope}
    \draw (0,1) -- (0,0) -- (1,-0.5) -- (2,0) -- (2,3); \draw (0,2) -- (0,3);
    \node at (0,1.5) {$U$};
    \draw (-0.5,1) rectangle (0.5,2);

    \draw (3.5,1.5) node[left]
    {$\stackrel{J}{\rightarrow}$};

    \draw (4,3) -- (4,0) -- (5,-0.5) -- (6,0) -- (6,1); \draw (6,2) -- (6,3);
    \node at (6.4, 1.5) {$K UK^{-1}$}; \draw (4.7,1) rectangle (8.1,2);

    \draw (9.4,1.1) node[left] {$=$};

    \draw (10,1) -- (10,0) -- (11,-0.5) -- (12,0) -- (12,3);
    \draw (10,2) -- (10,3);
    \node at (10,1.5) {$U^{\dag}$}; \draw (9.4,1) rectangle (10.6,2);

  \end{scope}
\end{tikzpicture}
$$

\paragraph{Thermofield double states}

Important examples of the state-operator map include thermofield double (TFD)
states used in the thermofield dynamics,
where a thermal density operator is mapped to
a state (thermofield double state) in the doubled Hilbert space. 
(For our purpose of studying (boundary) unitary operators,
there is generically no (local) Hamiltonian,
and hence there is no simple finite temperature thermofield double state.
Nevertheless, thermofield double states still
serve as a useful example to introduce and discuss the KMS condition.)
In TFD states, states from the first and second Hilbert spaces
are paired up by using energy eigenvalues:
\begin{align}
  \label{TFD type II}
  |\rho_{\epsilon} \rangle \!\rangle
  =
  ({1}/{\sqrt{Z}})
  \sum_{i}
  e^{ - \epsilon E_i}
  |E_i\rangle |E_i\rangle^*,
\end{align}
where $|E_i\rangle$ is the eigen state
of the Hamiltonian $H$ with energy $E_i$,
and
$|E_i\rangle^*$ is the time-reversal partner of $|E_i\rangle$,
satisfying  
$
(K H K^{-1}) |E_i \rangle^* = E_i |E_i \rangle^*. 
$
They evolve in time according to
$
+i \frac{d}{dt} |E_i(t) \rangle = H |E_i(t) \rangle
$
and 
$
-i \frac{d}{dt} |E_i(t)\rangle^* = K H K^{-1} |E(t) \rangle^*
$,
respectively.

%

The TFD state $|\rho_{\epsilon} \rangle \!\rangle$ is a purification
of the thermal density matrix $\rho_{\epsilon}= (1/Z) e^{-2 \epsilon H}$
at inverse temperature $2\epsilon$.
While
$|\rho_{\epsilon}\rangle\!\rangle$ is not maximally entangled
between the in and out Hilbert spaces for $\epsilon >0$,
it can be used as a reference state to 
invoke the operator-state map.
In the context of quantum many-body physics and quantum field theory,
the TFD state is a convenient reference state,
which provides a finite (but small) regularization (cutoff)
$\epsilon>0$.
For example,
the state corresponding to the unitary evolution operator
$U(t) = \exp( -i t H)$ is given by
$
| {U}_{\epsilon} (t) \rangle\!\rangle
= (U(t)\otimes I) | \rho_{\epsilon} \rangle\!\rangle
$.
Note that the state $| {U}_{\epsilon} (t) \rangle\!\rangle$
enjoys the shift property,
$
| {U}_{\epsilon} (t) \rangle\!\rangle
= (U(t)\otimes I) | \rho_{\epsilon} \rangle\!\rangle
=
(I\otimes KU(t)^{\dag}K^{-1}) | \rho_{\epsilon}\rangle\!\rangle
$.

In the context of TFD, the modular conjugation operator is
conventionally called the tilde conjugation
\cite{Ojima:1981ma}.
The antiunitary modular conjugation operator $J$ satisfies
\begin{gather}
  J^2 =1,
  \quad
  J|\rho_{\epsilon} \rangle\!\rangle
  =
  |\rho_{\epsilon} \rangle\!\rangle,
  \quad
  J \mathcal{A}_{{\it in}} J = \mathcal{A}_{{\it out}},
\end{gather}
where $\mathcal{A}_{{\it in}}/\mathcal{A}_{{\it out}}$
is the operator algebra
acting on $\mathcal{H}_{{\it in}}$/$\mathcal{H}_{{\it out}}$.
Observing that $|U_{\epsilon}\rangle\!\rangle$ is
stationary (invariant) under 
$
U(t) \otimes KU(t)K^{-1}
$,
we can introduce the modular Hamiltonian
by 
$\exp (- it \bar{H}) =
U(t) \otimes KU(t)K^{-1}$,
\begin{align}
  \bar{H}
  &= H\otimes I - I \otimes K H K^{-1}
    \nonumber \\
&=
                   H \otimes I - J (H\otimes I) J,
\end{align}
which generates a time-translation in the doubled Hilbert space.
The modular Hamiltonian satisfies 
\begin{align}
  &
  \bar{H} |\rho_{\epsilon} \rangle\!\rangle
  =
  0,
    \quad
   J \bar{H} J = - \bar{H},
    \quad
    \Delta^{-it} \mathcal{A}_{{\it in}/{\it out}} \Delta^{+it}
    =
    \mathcal{A}_{{\it in}/{\it out}}.
\end{align}
where $\Delta = \exp ( - \beta \bar{H})$ is the modular operator.

\subsection{The KMS condition}

Let us now review the KMS condition.
It is the statement characterizing
states (density matrices), and it reads:
\begin{align}
  \label{KMS II}
\langle  A B (t) \rangle_{\beta}
  =
 \langle B(t-i\beta) A  \rangle_{\beta}
\end{align}
for any two operators $A$ and $B$, where
$
  B(t)
  :=
    e^{i t H} B e^{- i t H},
$
and
$
  \langle  \cdots \rangle_{\beta}
  :=
    \mathrm{Tr}\,(\cdots e^{- \beta H})/\mathrm{Tr}\, (e^{-\beta H}).
$
\footnote{
For a finite-dimensional Hilbert space,
the KMS condition follows simply from the cyclic property of the trace,
$
  \mathrm{Tr}
  \left(
    e^{ - \beta H}
    A e^{ i t H} B e^{ - itH}
  \right)
  =
  \mathrm{Tr}
  \left(
    e^{ - \beta H}
    e^{ i (t- i\beta)H}
    B
    e^{ -i (t- i\beta)H}
    A
  \right).
$
However, the KMS condition holds beyond the finite Hilbert space setting.
Note that the expression using the trace
is meaningful only when operators such as the density matrix
$e^{- \beta H}$ belong to the trace class.
}
The KMS condition can be rephrased in the language of thermofield dynamics
\cite{Ojima:1981ma};
The KMS condition is nothing but the statement
\begin{align}
  \label{KMS I}
  J
  \Delta^{1/2}
  O | \rho_{\beta/2} \rangle\!\rangle
  =
  O^{\dag} | \rho_{\beta/2} \rangle \!\rangle,
  \quad
  O\in \mathcal{A}_{{\it out}}.
\end{align}
To see the connection,
we start from the TFD representation of
the correlator, 
$\langle AB(t) \rangle_{\beta}
=
\langle \! \langle
\rho_{\beta/2}| AB(t) | \rho_{\beta/2} \rangle\!\rangle$
(where on the RHS
we write $A\equiv A\otimes I, B(t)\equiv B(t)\otimes I$
by abusing notation).
Using \eqref{KMS I},
\begin{align}
  \label{derivation KMS 1}
  \langle AB(t) \rangle_{\beta}
  &=
    \left(
    J \Delta^{1/2}A | \rho_{\beta/2} \rangle\!\rangle,
    J \Delta^{1/2}B^{\dag}(t) | \rho_{\beta/2} \rangle\!\rangle
    \right)
    \nonumber \\
  &=
    \left(
     \Delta^{1/2}A | \rho_{\beta/2} \rangle\!\rangle,
     \Delta^{1/2}B^{\dag}(t) | \rho_{\beta/2} \rangle\!\rangle
    \right)^*
  \nonumber \\
  &=
  \langle \! \langle
  \rho_{\beta/2}| B(t) \Delta^{1/2}\cdot \Delta^{1/2}A | \rho_{\beta/2} \rangle\!\rangle,
\end{align}
where we use $(*, *)$ to represent the inner product in the doubled Hilbert
space,
and noted that $J$ is anti-unitary.
Since 
$|\rho_{\beta/2} \rangle\!\rangle
  =\Delta^{-1/2}|\rho_{\beta/2} \rangle\!\rangle$,
we conclude the KMS condition:
\begin{align}
  \langle  AB(t) \rangle_{\beta}
   &=
  \langle \! \langle
     \rho_{\beta/2}|
     \Delta^{-1/2}\Delta^{-1/2}
     B(t) \Delta A
    | \rho_{\beta/2} \rangle\!\rangle
     \nonumber \\
    &=
  \langle \! \langle
      \rho_{\beta/2}|
      \Delta^{-1}
      B(t) \Delta A
    | \rho_{\beta/2} \rangle\!\rangle
      \nonumber \\
  &=
\langle  B (t-i \beta)A \rangle_{\beta}.
\end{align}


The point is that the modular conjugation
operator $J$ effectively implements the cyclic property of the trace,
without relying on the finite dimensionality of the Hilbert space.
For our later applications,
what corresponds to \eqref{KMS I} is \eqref{J symmetry},
$J|U\rangle\!\rangle = |U^{\dag}\rangle\!\rangle$,
where the temperature is infinity.
At infinite temperature, 
the Schwinger-Keldysh trace satisfies
$
 \mathrm{Tr}\, [U^{\dag} V]
  =
  \mathrm{Tr}\,
  [V U^{\dag}]
$
for two unitary operators $U$ and $V$,
which is just the cyclicity of the trace.
In the state language, this follows from the existence of
modular conjugation operator.
Following \eqref{derivation KMS 1} with $\beta=0$,
\begin{align}
  \frac{1}{\mathcal{N}}
  \mathrm{Tr}
  \left[
  U^{\dag}V
  \right]
   &=
     \left(
    J U^{\dag}|\Omega \rangle\!\rangle, 
    J V^{\dag}|\Omega\rangle\!\rangle
    \right)
     \nonumber \\
    &=
     \left(
     U^{\dag}|\Omega\rangle\!\rangle, 
     V^{\dag}|\Omega \rangle\!\rangle 
    \right)^*
     \nonumber \\
   &=
     \left(
     V^{\dag}|\Omega\rangle\!\rangle, 
     U^{\dag}|\Omega \rangle\!\rangle 
    \right)
   =
     \frac{1}{\mathcal{N}}
     \mathrm{Tr}
\left[ 
     V U^{\dag}
 \right].
\end{align}

The KMS condition also follows from the shift property:
We note that the inner product
$
  \langle \!\langle U |V\rangle\!\rangle
  =
   {\mathcal{N}}^{-1}
    \sum_{i,j}
    \big(
    U|i\rangle \otimes K|i\rangle,
    V|j\rangle \otimes K|j\rangle
    \big)
  =
  {\mathcal{N}}^{-1}
  \mathrm{Tr}\,[ U^{\dag}V ]
$
can be computed by first using the shift property of $|\Omega\rangle\!\rangle$:
\begin{align}
  \langle \!\langle U |V\rangle\!\rangle
  &=
    \frac{1}{\mathcal{N}} 
    \sum_{i,j}
    (\langle i|)
    \otimes
    (\langle i|UK)
    (|j\rangle \otimes KV^{\dag}|j\rangle)
    \nonumber \\
  &=
    \frac{1}{\mathcal{N}} 
    \sum_{i}
    \Big(
    KU^{\dag}|i\rangle,
    KV^{\dag}|i\rangle
    \Big)
    \nonumber \\
  &=
     \frac{1}{\mathcal{N}} 
    \sum_{i}
    \Big(
    V^{\dag}|i\rangle,
    U^{\dag}|i\rangle
    \Big)
    =
    \frac{1}{\mathcal{N}}
    \mathrm{Tr}
    \left[
    VU^{\dag}
    \right].
\end{align}
Thus, the shift property
implies/is consistent with
the cyclicity of the trace:
$
\mathrm{Tr}\, [ U^{\dag} V]
=
\mathrm{Tr}\, [V U^{\dag}]
$.

\section{Gauging symmetries}
\label{Gauging symmetries}


\subsection{ Review: Gauging static (topological) phases}

Gauging a global symmetry is a useful framework to detect non-trivial (symmetry-protected) topological phases of matter
(see, for example,
\cite{2008PhRvB..78s5424Q,2013PhRvB..87o5114C, 2014arXiv1403.1467K,
  2014arXiv1404.6659K,
  2015JHEP...12..052K,
  2016RvMP...88c5005C}).
Here, by gauging, we mean introducing a non-dynamical, background gauge field
associated with the symmetry group.
In the following, our goal is to extend this paradigm to
unitary operators with symmetries; we will discuss the gauging procedure for topological/anomalous unitary operators. 

Let us first recall a few essential points of the gauging procedure for the case of static topological phases. 
To be concrete, suppose we have a static gapped (topological) phase described by 
the Euclidean path integral which is given schematically by
$
  Z[X] = \int \mathcal{D}\phi\,
  e^{- S[\phi,X]},
$
where $\phi$ symbolically represents the ``matter'' degrees of freedom,
and $S[\phi,X]$ is the Euclidean action on a closed $(d+1)$-dimensional spacetime manifold $X$.
In the presence of a background gauge field, we consider
\begin{align}
  Z[X,A] = \int \mathcal{D}\phi\,
  e^{- S[\phi,X, A]}.
\end{align}
(Here, for simplicity, we mainly focus on on-site unitary symmetry.
It is also possible to gauge spacetime symmetry, such as time-reversal,
reflection, and other space group symmetry, by considering, e.g., unoriented spacetime
\cite{2014arXiv1403.1467K, 2014arXiv1404.6659K, 2015JHEP...12..052K, 2014PhRvB..90p5134H}.)
For gapped phases (with the unique ground state),
the effective action $-\ln Z[X,A]$ is expected to be a local
functional of $A$. It may also have a pure imaginary, topological part,
signaling a non-trivial topological response of the ground state,
$Z[X,A]\sim \exp i S_{{\it top}}[X,A]$.
The topological term $S_{{\it top}}[X,A]$ can be thought of as a topological invariant characterizing the topological phase.

As an example, let us consider gapped phases in (1+1) spacetime dimensions, protected by on-site unitary symmetry.
We consider the Euclidean path integral on the spacetime torus $T^2$.
The non-trivial background gauge field configurations are then characterized by holonomies (Wilson loops) along the
two non-contractible loops on $T^2$.
The effect of the background can be thought of as twisting
boundary conditions of the matter field $\phi$ along the two non-contractible loops,
$\phi(\tau+T,x)= g\cdot\phi(\tau, x)$
and 
$\phi(\tau,x+L)= h\cdot\phi(\tau, x)$,
where $\tau\in[0,T]$ and $x\in [0,L]$ coordinatize the temporal and spatial directions, respectively,
and $g$ and $h$ are elements of the symmetry group.
We thus consider
\begin{align}
  \label{1+1d example}
  Z[T^2,(g,h)]
  &= \int\limits_{
    \substack{
  \phi(\tau+T,x)= g\cdot\phi(\tau, x),\\
  \phi(\tau,x+L)= h\cdot\phi(\tau, x)
  }}
  \mathcal{D}\phi\,
  e^{-S[\phi,T^2]}.
\end{align}
The topological term, i.e., the phase of the partition function,
is known to be classified by $H^2(G, U(1))$
\cite{1990CMaPh.129..393D,2013PhRvB..87o5114C,2017JHEP...04..100S}.
More generally, (especially in the case of orientation reversing symmetries), $S_{{\it top}}[X,A]$ can be thought of as a topological quantum field theory which depends only 
on the cobordism class of $[X, A]$~\cite{2014arXiv1403.1467K} (including spin structures in the case of fermions~\cite{2015JHEP...12..052K}), 
and is denoted by $\Omega_{d+1}^{\rm str}(B G)$. Here, ${\rm str}$ refers to the corresponding spin (or pin) structure for fermions and $B G$ is the classifying space of $G$.
When there is no symmetry, we simply put a single point as $BG$, 
$BG=pt$.
In this language, the topological term 
may be viewed  as a homomorphism 
$e^{i S_{top}} : \Omega_{d+1}^{\rm str}(B G) \to U(1)$.
Hence, the torsion part of the cobordism group ${\rm Tor}\ \Omega_{d+1}^{\rm str}(B G)$ can be used to provide a classification of topological phases protected by a symmetry group $G$~\cite{2014arXiv1403.1467K, 2015JHEP...12..052K, Freed2016}. For instance, time-reversal symmetric fermionic systems in (1+1) spacetime dimensions with $T^2=1$ have a $\Omega_{2}^{\rm Pin_-}(B G)=\mathbb{Z}_8 $ classification and the partition function on $\mathbb{R}P^2$ can be used as the corresponding $\mathbb{Z}_8$ topological invariant.

Quite often it is also possible to extract the topological term using the canonical (operator) formalism, in particular, 
solely from the ground state. 
The partition function \eqref{1+1d example} can be written in the operator formalism as
$
  Z[T^2, (g,h)]
  =
  \mathrm{Tr}_{h}\, \left[
  V_g\,
  e^{- T H_h}
  \right]$.
Here, $H_h$ is the system's Hamiltonian with twisted spatial boundary condition by $h$, and 
the trace is taken in the Hilbert space with the twisted boundary condition;
$V_g$ implements the symmetry operation $g$ in the ($h$-twisted) Hilbert space.
In the zero-temperature limit $T\to \infty$, the ground state dominates the partition sum,
\begin{align}
  \label{gauging static case}
  Z[T^2, (g,h)]
  =
  { }_h\langle  {\it GS}|\,
  V_g\,
  |  {\it GS}\rangle_h,
\end{align}
where $|{\it GS}\rangle_h$ is the ground state in the $h$-twisted sector.
Observe that the twisting boundary condition in the temporal
direction is implemented as the operator insertion $V_g$ within the trace.


Our strategy to study the anomalous properties of unitary operators
is to map them to corresponding states in the doubled Hilbert space (the operator-state map).
In particular, when the mapped states are short-range entangled,
which may be viewed as unique ground states of some gapped
Hamiltonians, we can use tools from the physics of
symmetry-protected topological phases to study the mapped states;
we can follow the gauging procedure outlined above for static topological phases of matter.
\footnote{
It is not entirely obvious for which Hamiltonian they are considered to be ground states. 
While not unique, such ``parent'' Hamiltonian can be constructed formally as
$
\mathbb{H} =
(U_{{\it out}}\otimes I_{{\it in}})
\mathbb{H}_0
(U^{\dag}_{{\it out}}\otimes I_{{\it in}})
$
where $\mathbb{H}_0$ is the gapped parent Hamiltonian for $|\Omega\rangle\!\rangle$.}
For the rest of this section, we will develop the gauging procedure
for unitary and anti-unitary symmetries, by
focusing first on
the ``temporal'' component of background gauge fields.
In particular,
we will observe that,
while time-reversal symmetry is antiunitary in
the original (single) Hilbert space,
it can be implemented as a unitary on-site
symmetry (the KMS symmetry),
and can be gauged following the standard procedure.
We will also establish
the connection between 
the temporal gauging procedure and 
the approach in Ref.\ \cite{2016PhRvB..93x5145V}
that deals with anomalous operator algebras appearing
on boundaries of 1d topological Floquet systems. 
In Sec.\ \ref{Spatial gauging},
we will also discuss spatial gauging
(turning on spatial components of background gauge fields)
--
the idea will be further developed
in the following sections by taking examples of various kinds.
(While we use the language of the operator-state map, and the doubled Hilbert
space, this may not be entirely necessary to develop the gauging procedure,
although we find it is quite convenient in many cases.
We will mention the perspective without using the operator-state map when possible.)

\subsection{Gauging unitary symmetries}
\label{Gauging unitary symmetries}

%

Let us consider a unitary time-evolution operator $U$ with symmetries.
We denote a symmetry group by $\mathscr{G}$.
For a given element $g \in \mathscr{G}$,
there is a unitary or an anti unitary operator $V_g$ acting
on the (physical) Hilbert space $\mathcal{H}$.
We say a unitary $U$ is symmetric under $\mathscr{G}$ when
\begin{align}
  \label{unitary sym}
  &
  V^{\ }_g U V_g^{-1} = e^{i \phi_g(U)} U,
    \quad
    V_g: \mbox{unitary}
   \\
  \label{a-unitary sym}
    &
  V^{\ }_g U V_g^{-1} = e^{i \phi_g(U)} U^{\dag},
      \quad
      V_g: \mbox{anti unitary}.
\end{align}
Here, note that we allow a projective phase in these operator algebras.
Such projective phases may appear when unitary operators are
realized on the boundary of topologically non-trivial bulk (Floquet) unitaries:
While symmetry can be realized in the bulk without projective phases, 
boundary unitaries can be anomalous and may pick up projective phases when acted by symmetries
\cite{2016PhRvB..93x5145V}. 
As we will see momentarily, the projected phases can be detected by
introducing a temporal component of the background gauge field.

Let us start with the case of unitary symmetry.
To discuss the gauging procedure, we begin by noting that
while symmetry $g$ acts on $U$ by conjugation, $U\to V^{\ }_gU V_g^{-1}$,
it acts on $|U\rangle\!\rangle$ as $
|U\rangle\!\rangle \to
 \left[
  V_g\otimes K V_g K^{-1}
  \right]
  | U \rangle\!\rangle
$.
Now, if we view $|U\rangle\!\rangle$ as a ground state (of a gapped parent Hamiltonian),
we consider, following the static case \eqref{gauging static case},
\begin{align}
  \label{part fn with twist in t direction}
  Z_g :=
  \langle \! \langle U|\,
  V_g \otimes K V_g K^{-1}\,
  |U\rangle\!\rangle.
\end{align}
This quantity can be interpreted as a partition function
in the spacetime manifold $S^1\times M$ (where $M$ is the spatial part)
in the presence of twisted boundary condition by $g$ in the temporal direction.
Here, ``time'' is a fictitious one,
and the time-evolution is generated by the putative parent Hamiltonian;
$V_g \otimes K V_g K^{-1}$ should be the symmetry of the parent Hamiltonian.
The phase of this partition function may detect an anomaly (topological information) of $|U\rangle\!\rangle$.
Using the shift property of $|U\rangle\!\rangle$,
\begin{align}
  V_g \otimes K V_g K^{-1}
  |U\rangle\!\rangle
  &=
  V_g U \otimes KV_gK^{-1}
  |\Omega \rangle\!\rangle
    \nonumber \\
   &=
  (V_g U V^{-1}_g \otimes I) 
  |\Omega \rangle\!\rangle
     \nonumber \\
  &=
    |V_g U V^{-1}_g \rangle\!\rangle,
\end{align}
so the ``partition function'' \eqref{part fn with twist in t direction}
is nothing but the overlap $
\langle \! \langle U | V_g U V^{-1}_g \rangle\!\rangle.
$
It can be further rewritten as
\begin{align}
  \label{twisted part fun, unitary}
  Z_g
   &=
     \mathcal{N}^{-1}
    \mathrm{Tr}\,
    \left[
    U^{\dag}V_g U V^{-1}_g 
    \right].
\end{align}
When $U$ is symmetric in the sense that $V_g U V^{-1}_g = e^{i \phi_g(U)} U$,
\begin{align}
  \label{trsf U under unitary sym}
    |V_g U V^{-1}_g \rangle\!\rangle
  =
    e^{ i \phi_g(U)}| U \rangle\!\rangle,
\end{align}
and the partition function is a pure phase quantity, $Z_g = e^{i \phi_g(U)}$.
The non-zero phase signals the anomalous nature of the unitary operator.
Note also that by construction, $|I\rangle\!\rangle=|\Omega\rangle\!\rangle$ is invariant under $V_g \otimes K V_g K^{-1}$.

As mentioned around \eqref{overlap = SK},
we can also interpret $Z_g$ in terms of the Schwinger-Keldysh path-integral (trace) with a temporal background gauge field.

\subsection{Gauging the KMS symmetry}
\label{Gauging the KMS symmetry}

Let us now turn to the case of antiunitary symmetry.
To be concrete, we will work with a time-reversal symmetric unitary,
\begin{align}
  \label{t sym}
&
T U T^{-1} = e^{i \phi_T (U)} U^{\dag}.
\end{align}
We note that in general $T$ can be written as  
$T = W\times (\mbox{complex conjugation})$ where $W$ is a unitary matrix:
In the basis $\{ |i\rangle\}$, $T$ is defined by its action on
$\{|i\rangle\}$ as
\begin{align}
  \label{T sym, general}
  T |i\rangle =
  \sum_{j}
  W_{ij}|j\rangle,
  \quad
  Ti T^{-1} = -i,
\end{align}
with $W_{ij} = \langle j|W|i \rangle$.
We note that the fact that
time-reversal squares to the identity,
$T^2=I$,
possibly up to the fermion number parity operator
for fermionic systems,
$T^2= (-1)^F$, 
imposes a restriction on
the projective phase \eqref{t sym}.
To see this, we first find 
the hermitian conjugate of  \eqref{t sym},
$
T U^{\dag} T^{-1} = e^{ - i \phi_T(U)} U$, and then apply $T$,
which gives
$T^2 U^{\dag} T^{-2} = e^{i \phi_T(U)} T U T^{-1} = e^{2i \phi_T(U)} U^\dag$.
Assuming $U$ is fermion number parity even (odd),
the projective phase is quantized as
$e^{2 i\phi_T(U)}=\pm 1$.

To gauge time-reversal symmetry,
we first need to discuss how time-reversal acts in the doubled Hilbert space,
as we did for the case of unitary symmetry.
We should note that antiunitary symmetry does not allow
tensor factorization in the doubled Hilbert space,
in contrast with unitary symmetry $g$,
which acts on the doubled Hilbert space as $V_g \otimes K V_g K^{-1}$.
Nevertheless,
the time-reversal $T$ can be naturally 
extended to the doubled Hilbert space as
\begin{align}
  \label{T sym, extended}
  T| i\rangle |j\rangle^*
  &=
  \sum_{i'j'}
  W^{\ }_{ii'} (W^{\dag})_{j'j} |i'\rangle |j^{\prime}\rangle^*
    \nonumber \\
  &=
  \sum_{i'j'}
  W^{\ }_{ii'} (W^{\ }_{jj'})^* |i'\rangle |j^{\prime}\rangle^*.
\end{align}
Here we recall that $\{| i\rangle^*\}$ is the conjugate
representation of $\{| i \rangle\}$.
$T$ is an antiunitary operator on
$\mathcal{H}_{{\it out}}\otimes \mathcal{H}_{{\it in}}$.

One can check easily 
$T|\Omega \rangle\!\rangle = |\Omega\rangle\!\rangle$.
The symmetry condition $T U T^{-1}=
e^{i \phi_T(U)} U^{\dag}$ is translated into
$T|U\rangle\!\rangle =
| T U T^{-1}\rangle\!\rangle
=
e^{i \phi_T(U)} 
|U^{\dag}\rangle\!\rangle$
(c.f., \eqref{trsf U under unitary sym}).
Then, 
analogously to \eqref{twisted part fun, unitary},
we can consider the overlap
\begin{align}
  \label{overlap, T symmetry}
  \langle\!\langle U^{\dag}|T| U \rangle\!\rangle
  &=
  \langle\!\langle U^{\dag}| TUT^{-1} \rangle\!\rangle
    \nonumber \\
  &=
  \mathcal{N}^{-1}
  \mathrm{Tr}
  \left[ U T U T^{-1} \right]
  =
  e^{i \phi_T(U)}.
\end{align}
As in \eqref{part fn with twist in t direction} 
this overlap can be interpreted
as the partition function on $S^1\times M$
with twisted temporal boundary condition
by some symmetry.
The relevant symmetry is $TJ$ -- 
the composition of time-reversal and
modular conjugation
--
which we will call the KMS symmetry.
This symmetry is unitary, while
both $J$ and $T$ are antiunitary. 

To see this,
we can first verify that the combined operation $TJ$
is a symmetry of $|U\rangle\!\rangle$,
\begin{align}
  \label{TJ symmetry of U}
  TJ |U\rangle\!\rangle
  &=
  T (U^{\dag}\otimes I) |\Omega \rangle\!\rangle
  =
  (TU^{\dag}T^{-1}\otimes I) T|\Omega \rangle\!\rangle
  \nonumber \\
  &
    =
    e^{ -i \phi_T(U)}
  (U\otimes I) |\Omega \rangle\!\rangle
  =
    e^{ -i \phi_T(U)}
  |U\rangle\!\rangle,
\end{align}
where we recall that $J A |\Omega \rangle\!\rangle = A^{\dag}|\Omega \rangle\!\rangle$.
Namely, neither $J$ nor $T$ are a symmetry in the doubled Hilbert space
(they do not leave $|U\rangle\!\rangle$ invariant),
but $JT$ is ($JT$ leaves $|U\rangle\!\rangle$ invariant
up to possibly a phase factor $e^{ -i \phi_T(U)}$).
In other words, the KMS condition,
once combined with time-reversal, 
can be ``promoted'' to a unitary symmetry
in the doubled Hilbert space.
The KMS symmetry, here identified by using the operator-state map,
also has its counterpart in the Schwinger-Keldysh path integral language.
In the path-integral language, Ref.\ \cite{2015PhRvB..92m4307S} (see also \cite{Crossley:2015evo})
proposed a symmetry of the Schwinger-Keldysh path integral under 
$\psi_{\sigma}(t, \mathbf{r}) 
\to 
\psi^*_{\sigma}(-t + i\sigma \beta/2,\mathbf{r})$,
$\psi^*_{\sigma}(t, \mathbf{r}) 
\to 
\psi_{\sigma}(-t + i\sigma \beta/2,\mathbf{r})$,
as the KMS condition. Here, $\psi_{\sigma}(t,\mathbf{r})$ schematically represents
quantum fields in the Schwinger-Keldysh path integral
where $\sigma=\pm$ represents the forward and backward branches.
Note that the KMS symmetry can be defined 
(and gauged) at finite temperature,
although in this paper we set temperature to be infinite.

%
%
%

Now, the KMS symmetry, being unitary on-site symmetry in the doubled Hilbert space, 
can be gauged in a straight forward way.
Following the static case
\eqref{gauging static case}, we consider 
the partition function with twisted temporal boundary condition
by the KMS symmetry,
\begin{align}
  Z_{{\it KMS}}
  =
  \langle \! \langle U| (TJ)|U \rangle\!\rangle.
\end{align}
Using \eqref{TJ symmetry of U}
$Z_{{\it KMS}}$
is nothing but (the complex conjugate of)
\eqref{overlap, T symmetry},
\begin{align}
  Z_{{\it KMS}}
  =
  \langle \! \langle U| TU^{\dag}T^{-1} \rangle\!\rangle
  =
  e^{ -i \phi_T(U)}.
  \end{align}

\subsection{Unitarity condition and chiral symmetry in the doubled Hilbert space}
\label{Unitarity condition and chiral symmetry in the doubled Hilbert space}


In the forthcoming sections, we will study
the anomalous properties of unitary operators 
using the gauging procedure outlined above.
It should be noted 
however that it is not entirely obvious
if all anomalous (topological) properties of
unitaries can be detected this way.
%
For example, 
we should note that
the state-operator map can be applied to any operator
acting on the original Hilbert space, not just unitaries.
Hence, we need to narrow our focus down to
the set of states in the doubled Hilbert
space that correspond to unitary operators in the original Hilbert space.
\footnote{
To illustrate this point, 
let us consider the Berry phase of mapped states in the doubled Hilbert space,
when
we have 
unitaries $|U(R)\rangle\!\rangle$
parameterized by adiabatic parameters $R=(R_1, R_2,\cdots)$. 
Noting that the Berry connection is given explicitly by
$
A^i = 
i
\langle \! \langle
U|
(\partial/\partial R_i)
|U \rangle\!\rangle
=
i
\mathrm{Tr}\,
[
U^{\dag} (\partial U/\partial R_i)
]
$,
the Berry phase associated to any closed loop in the parameter space
is quantized to an integer multiple of $2\pi$, 
$
  \oint A^idR_i = 2\pi \times {\it integer}.
$
Clearly, this is not the case for generic states in ${\cal H}_{{\it out}}
\otimes {\cal H}_{{\it in}}$.
This is one of the consequences of the unitarity condition.}

As an illustration, let us consider one of the simplest examples,
Floquet unitaries in one spatial dimension
with on-site unitary $\mathbb{Z}_2$ symmetry.
Such unitaries are known to be classified by $\mathbb{Z}_2$
\cite{2016PhRvB..93x5145V}.
On the other hand, once such unitaries are mapped to states,
we are to consider short-range entangled states
with on-site unitary $\mathbb{Z}_2$ symmetry.
Since $H^2(\mathbb{Z}_2,U(1))=0$, there is no non-trivial topological phase.
This disagreement presumably comes from the fact that 
the set of short-range entangled states (with $\mathbb{Z}_2$ symmetry)
in the doubled Hilbert space includes states which do not correspond to unitaries.

For matrix product unitaries,
the unitarity condition (requirement) can be 
taken into account by using the standard form of matrix product unitaries
\cite{2017JSMTE..08.3105C,2018PhRvB..98x5122S}.
Moreover, 
the chiral unitary index (GNVW index),
a rational number that characterizes 
asymmetric quantum information flow,
can be introduced to classify unitaries.
\cite{2017JSMTE..08.3105C,2018PhRvB..98x5122S,2017arXiv170307360F,2017PhRvL.118k5301H}.
For the case of non-interacting fermionic systems (Gaussian unitaries),
we can impose an additional symmetry,
the so-called chiral symmetry,
in the doubled Hilbert space,
to limit our focus to states corresponding to unitary operators
(and enforce the quantization of the Berry phase)
\cite{2017PhRvB..96o5118R,2017PhRvB..96s5303Y,2019arXiv190803217G}.
(In the context of free fermion systems (Gaussian unitaries),
the operator-state map is called the hermitian map.)

In the following, we will deal with 1d examples with time-reversal symmetry,
for which the chiral unitary index vanishes.
Following the case of on-site unitary symmetries for (bosonic) 1d unitaries
\cite{2017JSMTE..08.3105C,2018arXiv181209183G},
we expect that the anomalies (group cohomology class) 
associated with the KMS symmetry (together with other symmetries)
are enough to classify these unitaries.



%

\subsection{Spatial gauging}
\label{Spatial gauging}

The spatial component of the background gauge field can also implemented in the unitary operator.
For example, the spatial component of the background KMS gauge field can be introduced by twisting the spatial boundary condition. 
To do this, we need to have a closer look at
the local (spatial) structure of unitaries.
As we will discuss in the next section,
once a unitary is given as a matrix product unitary, 
the spatial component of the background gauge field
can be introduced, following the gauging procedure of matrix product states  
\cite{2017JHEP...04..100S}.
(See below around \eqref{gauing KMS in MPU}).
Another way to introduce spatial gauging is to make use of
a parent Hamiltonian that has $|U\rangle\!\rangle$ as its ground state.
If it exists, we can introduce the background gauge field
by minimally coupling it to
matter degrees of freedom in the parent Hamiltonian.
We will discuss this in the forthcoming sections by using examples,
see Secs.\ \ref{Majorana fermion models}
and \ref{Anomalous unitary operators with $U(1)$ and discrete symmetries}.
Finally, we also note that 
it is known that 
torus partition functions with twisted boundary conditions
(topological invariants)
can be computed solely by using ground state wave functions
(without using Hamiltonians)
by using the partial swap operator
\cite{2012PhRvL.109e0402H, 2017JHEP...04..100S}.

In the presence of a spatial component of a gauge field,
the operator algebra \eqref{unitary sym}
can be generalized as
\begin{align}
    V^{\ }_g U(A_h) V_g^{-1} = e^{i \phi_g(U(A_h))} U(A_h)^{s(g)},
\end{align}
where $s(g)=1$ or $s(g)=-1=\dag$
when $V_g$ is a unitary or anti-unitary symmetry, respectively,
and $A_h$ is the background $h$ gauge field.
(Here, we are assuming $g$ is a non-spatial symmetry.
When $g$ is a spatial symmetry, e.g., parity, the gauge field
$A_h$ also has to be transformed -- see
\eqref{UV comm with A}.)   
Correspondingly, we can consider
the overlap
\begin{align}
  \label{part fn with twist in t direction with Ah}
  Z_g(A_h) &:=
  \langle \! \langle U(A_h)^{s_g}|\,
  V_g \otimes K V_g K^{-1}\,
  |U(A_h)\rangle\!\rangle
             \nonumber \\
  &=
    \mathcal{N}^{-1}
    \mathrm{Tr}\,
    \left[
    (U^{\dag}(A_h))^{s_g}V_g U(A_h) V^{-1}_g 
    \right],
\end{align}
which can be interpreted as a
partition function on $S^1\times M$
with
twisted temporal boundary condition by $g$,
and spatial background gauge field $A_h$ on $M$.

\section{Matrix product unitaries}
\label{Matrix product unitaries}

All locality-preserving 1d unitaries
(in bosonic systems)
can be represented in the form of a matrix product unitary
\cite{2017JSMTE..08.3105C,2018PhRvB..98x5122S}.
In this section, we discuss how we can gauge the KMS symmetry
in matrix product unitaries.
A matrix product unitary $U$
is expressed as
\begin{align}
  U
  =
  \sum_{\{i,j\}}
  \mathrm{Tr}
  \left(
  A^{i_1 j_1}  \cdots A^{i_L j_L} \right)
  \ket{i_1 \cdots i_L}
  \bra{j_1 \cdots j_L},
\end{align}
where $\{|i_1 \cdots i_L\rangle\}$
is the basis of the total Hilbert space of
the 1d chain consisting of $L$ sites,
given as a tensor product of basis states
$\{|i\rangle\}$
of the local Hilbert space at each site; 
$A$ is a $\chi\times \chi$ dimensional matrix
where $\chi$ is the ``bond-dimension'' of the auxiliary space.
By the operator-state map, 
the corresponding state in the doubled Hilbert space is
\begin{align}
  \label{MPU state}
  |U\rangle\!\rangle
  =
  \sum_{\{i,j\}}
  \mathrm{Tr}
  \left(
  A^{i_1 j_1}  \cdots A^{i_L j_L} \right)
  \ket{i_1 \cdots i_L}
  \ket{j_1 \cdots j_L}^*.
\end{align}
Once written in this form,
we can apply results from
matrix product states, in particular
their classification.
However, this does not fully capture
the full classification of unitaries.
The reason is that we have not included
the unitarity requirement, and the chiral unitary index.
(See Sec.\ \ref{Unitarity condition and chiral symmetry in the doubled Hilbert space}.)
References
\cite{2017JSMTE..08.3105C,2018PhRvB..98x5122S}
introduced the standard form of matrix product unitaries
that takes into account the unitarity requirement,
and defined the chiral unitary index. 
Using the standard form, 
symmetry protected indices can also be introduced for on-site unitary symmetry
\cite{2018arXiv181209183G}.
The gauging procedure we introduced
in the previous section is agnostic 
about the unitarity condition,
and hence, in particular, cannot capture
the chiral unitary index.
We however note that 
for unitaries of our interest,
namely, those that are invariant under time-reversal,
the chiral unitary index always vanishes.
At this stage, it is unclear if the gauging procedure misses
other topological/anomalous aspects of unitary operators.
Nevertheless, topological invariants (quantum anomalies)
derived from the gauging procedure 
provides a bona fide diagnostic of anomalous unitary operators. 


Let us now assume the unitary $U$ is time-reversal symmetric in the sense
that $T U T^{-1} = U^{\dag}$ (up to a projective phase)
where $T$ is time-reversal,
which, as in \eqref{T sym, general},
can be written as $T = WK$ with some unitary $W$.
The time-reversal $T$ can be naturally 
extended to the doubled Hilbert space as
in \eqref{T sym, extended}.
Together with the antiunitary modular conjugation operator,
$
  J( |i\rangle | j\rangle^*)
  =
  |j\rangle |i\rangle^*,
$
$J|U\rangle\!\rangle = |U^{\dag}\rangle\!\rangle$,
we can construct the KMS symmetry $JT$,
which is a unitary, on-site, $\mathbb{Z}_2$ symmetry.
For the matrix product unitary,
$JT$ acts on $|U\rangle\!\rangle$ as
\begin{align}
  &
  JT|U\rangle\!\rangle
  =
    \sum_{\{i,j,i',j'\}}
    \mathrm{Tr}\,
    \left(
    A^{j_1 i_1} \cdots A^{j_L i_L} \right)
    \nonumber \\
   &\quad
     \times
    W_{i_1 i'_1} (W_{j_1 j'_1})^*
    \cdots 
    W_{i_L i'_L} (W_{j_L j'_L})^*
    \ket{i'_1 \cdots i'_L}
    \ket{j'_1 \cdots j'_L}^*.
\end{align}
The invariance under $JT$ implies that
each matrix $A$ transforms as
\cite{2011PhRvB..83c5107C,
  2011PhRvB..84p5139S,2012PhRvB..85g5125P}
\begin{align}
  \label{JT projective}
  \sum_{ij}
  A^{j i}_{ab} W_{i i'} (W_{j j'})^*
  =
  e^{i\theta}
  \sum_{a'b'}
  (M^{\dag})_{aa'}
  A^{i'j'}_{a'b'}
  M_{b'b}
\end{align}
with some matrix $M$ and phase $e^{i\theta}$.
In one spatial dimension,
a unitary on-site $\mathbb{Z}_2$ symmetry alone does not
lead to non-trivial SPT 
phases, as $H^2(\mathbb{Z}_2, U(1))=0$.
(This is consistent with \cite{2017JSMTE..08.3105C}.)
\footnote{
  Note that in Ref.\ \cite{2017JSMTE..08.3105C}
  unitaries satisfying $U=U^{\dag}$ are called ``time-reversal symmetric''.
  Here, we stick with time-reversal which is realized as
  a antiunitary operation in the physical Hilbert space,
  as Wigner's symmetry representation theorem.
  Ref.\ \cite{2017JSMTE..08.3105C}
  also studied unitaries satisfying $U=U^T$, which
  corresponds (up to possibly a unitary operation) to our definition
  of time-reversal symmetry.
  For the latter case, 
  Ref.\ \cite{2017JSMTE..08.3105C} showed that there is no non-trivial unitary,
  consistent with $H^2(\mathbb{Z}_2, U(1))=0$.
}
However, in the presence of other symmetries,
we can discuss the discrete torsion phase
with KMS symmetry.

We can gauge the state $|U\rangle\!\rangle$ by the KMS symmetry;
the state under the twisted boundary condition by the KMS symmetry
is given by \cite{2019PhRvL.123f6403H}
\begin{align}
  \label{gauing KMS in MPU}
  |U\rangle\!\rangle_{{\it KMS}}
  &=
  \sum_{\{i,j\}}
  \mathrm{Tr}
  \left(
  A^{i_1 j_1}  \cdots A^{i_L j_L} M\right)
  \ket{i_1 \cdots i_L}
  \ket{j_1 \cdots j_L}^*.
\end{align}
Let us now imagine that 
$U$ is symmetric under an additional unitary symmetry $X$,
$X U X^{-1}=U$ (up to possibly a phase factor).
The torus partition function \eqref{gauging static case}
can be computed,
in the presence of another symmetry generator $X$,
${ }_{{\it KMS}}\langle \! \langle U| X |U\rangle\!\rangle_{{\it KMS}}$,
which extract a topological invariant (cocycle).
Note that once the matrix product operator form is given,
it is not necessary to use the parent Hamiltonian to gauge symmetries.

Using the operator-state map, we can map
$|U\rangle\!\rangle_{{\it KMS}}$ back to an operator $U_{{\it KMS}}$,
\begin{align}
    U_{{\it KMS}}
    &=
      \sum_{\{i,j\}}
      \mathrm{Tr}
      \left(
      A^{i_1 j_1}  \cdots A^{i_L j_L} M\right)
      \ket{i_1 \cdots i_L}
      \bra{j_1 \cdots j_L}.
\end{align}
This can be thought of as the gauged unitary operator,
in the presence of background KMS gauge field. 
When $U$ is symmetric under an additional unitary
on-site symmetry $X$,
$X$ induces an action on the auxiliary space
by a unitary matrix $M_X$, as in \eqref{JT projective}.
Then, the operator algebra between $X$ and $U_{{\it KMS}}$
is given by
\begin{align}
  &
  X\, U_{{\it KMS}}\, X^{-1}
    =
    e^{i \phi_{X, {\it KMS}}}\, U_{{\it KMS}},
\end{align}
where we note that
\begin{align}
  X\, U_{{\it KMS}}\, X^{-1}
  &=
    \sum_{\{i,j\}}
    \mathrm{Tr}
    \left(
    A^{i_1 j_1}  \cdots A^{i_L j_L} M^{\ }_X M M^{\dag}_X\right)
    \nonumber \\
  &\quad
    \times 
    \ket{i_1 \cdots i_L}
    \bra{j_1 \cdots j_L}
\end{align}
and 
$e^{i \phi_{X, {\it KMS}}}$
is the group cohomology phase,
$M_X M  = e^{i \phi_{X, {\it KMS}}}M M_X$.
Thus, (the phase of) the torus partition function
and the anomalous phase that appears in the 
operator algebra between the {\it gauged} unitary operator
and symmetry generator
is equivalent.

\paragraph{Example: the CZX model.}
As a simple example, let us consider the CZX model
\cite{2011PhRvB..84w5141C,2017JSMTE..08.3105C}.
It is defined on a one-dimensional lattice with
two-dimensional local Hilbert space at each site,
$\{\ket{0}, \ket{1}\}$.
The explicit matrix product unitary form is given as
\begin{align}
  A^{01} = \ket{0}\bra{+},
  \quad
  A^{10} = \ket{1}\bra{-},
  \quad
  A^{00}=A^{11}=0,
\end{align}
with two-dimensional internal (auxiliary) Hilbert space, and
$|\pm \rangle = |0\rangle \pm |1\rangle$.
The chiral unitary index is trivial for the CZX model.
The CZX unitary $U$ is invariant under time-reversal $K U K^{-1} =  U^{\dag}$.
Hence, under $JT$, $A$'s are transformed as
$A^{j i}_{ab} = e^{i\theta} (W^{\dag})_{aa'} A^{ij}_{a'b'} W_{b'b}$.
It is easy to check that we can take $W=\sigma_y$,
$A^{ij} = - \sigma_y A^{ji} \sigma_y^{\dag}$.

Now, let us consider an additional $\mathbb{Z}_2$ symmetry.
We can consider, for example, 
$X = \prod^{odd}_i s^x_{i}$,
which commutes with time-reversal.
(Here, $s^x_i$ is  the $x$-component of  a physical spin $1/2$ operator 
at site $i$.)
It is convenient to ``block'', i.e.,
take two adjacent spins as a single degrees of freedom;
at each site, we now have a four-dimensional local Hilbert space
spanned by
$\{\ket{00}, \ket{01}, \ket{10}, \ket{11}\}$.
Under blocking, we consider  
the matrix product unitary with
\begin{align}
  A^{0101}
  &=
    A^{01}A^{01}
    = \ket{0}\bra{+},
    \quad
  A^{0110}
  =
    A^{01}A^{10}
    = \ket{0}\bra{-},
    \nonumber \\
  A^{1001}
  &=
    A^{10}A^{01}
    = \ket{1}\bra{+},
    \quad
  A^{1010}
  =
    A^{10}A^{10}
    = -\ket{1}\bra{-}.
\end{align}
Under symmetry $X$, $A^{ijkl}\to A^{\bar{i}\bar{j}kl}$
(where $\bar{0}=1$ and $\bar{1}=0$).
One can check that
the invariance under $X$ can be implemented by
$A^{ijkl}
  \to A^{\bar{i}\bar{j}kl}
  =
  \sigma_x A^{ijkl} \sigma^{\dag}_x.
$
Now, while the $JT$ and $X$ commute
when acting on the physical Hilbert space,
in the two-dimensional auxiliary space,
$\sigma_y \sigma_x = -\sigma_x \sigma_y$,
implying that the CZX model is protected by time-reversal and $X$.

%
%
%
%
%
%

\section{Majorana fermion models}
\label{Majorana fermion models}

In this section, we consider unitary time-evolution operators
in Majorana fermion systems in one spatial dimension without/with time-reversal symmetry.
As a specific model, we consider the boundary unitaries which are realized
at the boundary of topological Floquet drives without/with time-reversal symmetry.
We first consider the model without time-reversal symmetry
(``the single copy theory'')
on the boundary of the 2d topological chiral Floquet drive
considered in \cite{2017arXiv170307360F}.
The time-reversal symmetric model (``the two copy theory'')
can then be constructed
from two copies of the above model with opposite chiralities.
We will then discuss non-trivial 2d time-reversal symmetric unitaries
that can be realized at the boundary of 3d topological Floquet systems.

\subsection{The single copy theory}

Let us first have a closer look at the single copy theory.
At the boundary of 2d topological chiral Floquet drive
\cite{2017arXiv170307360F},
discrete time-evolution
is given by a boundary unitary $S$,
which is a lattice translation operator (or shift operator):
\begin{align}
  S \lambda_{x} S^{\dag} = \lambda_{x+1}.
  \label{bdy unitary, single}
\end{align}
Here $\{\lambda_x\}$ is the set Majorana fermion operators defined on sites $x$
located at the boundary of the 2d system,
$\{\lambda_x, \lambda_{x'}\}= 2\delta_{xx'}$.
Throughout this section,
we impose the periodic boundary condition.
The translation operator $S$ can be written down explicitly as
\cite{2017arXiv171201148S}
\begin{align}
  S:=
  e^{i \delta}
  \lambda_1
  \frac{1+\lambda_1 \lambda_{2}}{\sqrt{2}}
  \cdots
  \frac{1+\lambda_{L-1} \lambda_{L}}{\sqrt{2}}.
\end{align}
The phase can be chosen such that the translation operator $S$ satisfies $S^L=1$
where $L$ is the total number of sites.
The phase factor satisfies
$e^{i\delta}=1$ when $L/2=4,5,8,9$ while $e^{i\delta}= e^{i\pi L}$ when $L/2=2,3,6,7$ (mod 8).

This shift operator is characterized by
non-zero chiral unitary index (GNVW index)
\cite{2017arXiv170307360F,2017PhRvL.118k5301H}.
The chiral unitary index can be defined without referencing to any symmetry,
and hence the topological Floquet drive does not require any symmetry for
its stability/existence.
As mentioned briefly in Sec.\ \ref{Unitarity condition and chiral symmetry in the doubled Hilbert space},
for Gaussian unitaries,
we can impose chiral symmetry to discuss the chiral unitary index.
This puts the system in symmetry class BDI (in the doubled Hilbert space)
-- see around \eqref{JR}.

As we will see, to discuss the chiral unitary index, we can impose
the unitarity condition on states in the doubled Hilbert space.

Beside the chiral unitary index,
we can also discuss an anomaly associated with 
the fermion number parity,
$
  (-1)^F = \prod_{n=1}^{L/2} (i \lambda_{2n-1} \lambda_{2n}).
$
(With the conservation of the fermion number parity,
the relevant Altland-Zirnbauer symmetry class is class D.)
We can verify that
the shift operator is odd under the fermion number parity
\cite{2016PhRvL.117p6802H},
\begin{align}
(-1)^F S (-1)^F = -S,
\label{eqn:podd}
\end{align}
and hence,
the partition function twisted by the fermion number parity is
\begin{align}
  \mathcal{N}^{-1}
  \mathrm{Tr}\,
  \left[
  (-1)^F S (-1)^F S^{\dag}
  \right]
  = (-1).
\end{align}
The $\mathbb{Z}_2$ phase (minus sign) on the RHS is indicative of a $\mathbb{Z}_2$ quantum anomaly, occurring at the boundary of the bulk 2d Floquet system.
The $\mathbb{Z}_2$ anomaly is 
independent of the chiral unitary
index, and provides an additional characterization.


Let us now have a closer look at how the operator-state map works in this problem.
We will be slightly generic and consider an arbitrary Gaussian unitary operator $U$.
It transforms Majorana fermion operators $\{ \lambda_a \}$ (satisfying $\{ \lambda_a, \lambda_b \} = 2\delta_{ab}$) as
\begin{align}
  U \lambda_a U^{\dagger} & = {\cal Q}_{ab} \lambda_b
\end{align}
where ${\cal Q}$ is a real orthogonal matrix.
To deploy the state operator map,
we introduce the doubled Hilbert space by considering 
the two sets of Majorana fermion operators
$\{\lambda_{i,x}\}$ and $\{\lambda_{o,x}\}$
acting on the in and out Hilbert spaces, respectively.
The construction of the reference state
proceeds in a way slightly different than the 
bosonic case reviewed in
Sec.\ \ref{The operator-state map and KMS condition}.
As the reference state \eqref{ref state},
we need to look for a maximally entangled state
in the ($\mathbb{Z}_2$-graded) fermionic Hilbert space,
which satisfies the shift property, and is invariant under
a properly defined modular conjugation operator. 
Conveniently,
the reference state can be taken as a ground state of the parent Hamiltonian
\begin{align}
  \mathbb{H}_0 = {i} \sum_x \lambda_{i,x} \lambda_{o,x}.
\end{align}
We identify the modular conjugation operator as\footnote{
  Generically, the modular conjugation operator
  should satisfy 
  $[J\mathcal{A}_{{\it in}}J, \mathcal{A}_{{\it in}}]=0$,
  while for the $J$ operator defined here,
  $J\lambda_i J$ and $\lambda_i$ anticommute.
  The $J$ operator here is actually the tilde conjugation
  in the thermofield dynamics \cite{Umezawa:1993yq}.
  While
  for bosonic systems
    the modular conjugation and 
    the tilde conjugation are equivalent,
    for ferminoic systems, 
    they differ by a Klein factor (Jordan-Wigner string)
    \cite{Ojima:1981ma}.}
\begin{align}
  J \lambda_{i,x} J^{-1} = \lambda_{o,x},
  \quad 
  J \lambda_{o,x} J^{-1} = \lambda_{i,x}.
\end{align}
    
One can check easily that
$J \mathbb{H}_0 J^{-1}= \mathbb{H}_0$ and hence $J|\Omega \rangle\!\rangle \equiv |\Omega\rangle\!\rangle$.
We consider the state $|U\rangle\!\rangle
= (U_o \otimes I_i) |\Omega \rangle\!\rangle$,
which can be thought of as a ground state of 
\begin{align}
  \mathbb{H} ={i} \sum_{xy} \lambda_{i, x} {\cal Q}_{xy} \lambda_{o, y}.
\end{align}
The shift property of $|\Omega\rangle\!\rangle$ can be read off from
$\mathbb{H}$ as
\begin{align}
  \mathbb{H}=
  {i}
\sum_{xy} \lambda_{i, x} ({\cal Q}_{xy} \lambda_{o, y})
=
  i
\sum_{xy}  ({\cal Q}^{-1}_{yx} \lambda_{i, x}) \lambda_{o, y}
\end{align}
where we noted $\mathcal{Q}_{yx}=\mathcal{Q}^{-1}_{xy}$. Hence, $|U\rangle\!\rangle = 
(U_o \otimes I_i) |\Omega\rangle\!\rangle
\propto
(I_{o} \otimes U^{\dag}_i)|\Omega \rangle\!\rangle
$.
We observe that $\mathbb{H}$ is not invariant under $J$,
while $\mathbb{H}_0$ is, as expected.

Let us now consider the unitary in \eqref{bdy unitary, single}.
Then, the parent Hamiltonian is 
\begin{align}
  \mathbb{H} = {i} \sum_x \lambda_{i, x} \lambda_{o, x+1}.
\end{align}
This is essentially the Hamiltonian of the Kitaev chain in its topologically non-trivial phase. 
The ground state is characterized by the $\mathbb{Z}_2$ topological invariant of
symmetry class D in one spatial dimension,
consistent with the $\mathbb{Z}_2$ anomaly \eqref{eqn:podd}.

On the other hand, the topological classification of 2d Majorana
Floquet drives is $\mathbb{Z}$ for the non-interacting case.
For interacting case, the (fermionic version of) chiral unitary index classifies
gapped (many-body localized) Floquet derives. 
Either way, the $\mathbb{Z}_2$ topological invariant of symmetry class D seems not to match with these classifications.
As mentioned in the previous section,
the key to realize is that there is more than class D symmetry,
which arises because of the doubling.
While $\mathbb{H}$ is not invariant under $J$,
$\mathbb{H}$ is invariant under the combination of $J$ and swap $R$:
\begin{align}
  \label{JR}
  (JR)\lambda_{i,x} (JR)^{-1} = \lambda_{i,x},
  \quad
  (JR)\lambda_{o,x} (JR)^{-1} = (-1)\lambda_{o,x},
\end{align}
where 
$R \lambda_{i,x} R^{-1} = \lambda_{o,x},
R \lambda_{o,x} R^{-1} = (-1)\lambda_{i,x}$.
Since $JR$ is antiunitary and $(JR)^2=1$,
imposing this symmetry puts the parent Hamiltonian in symmetry class BDI.
At least at the non-interacting level,
we then reproduce the known $\mathbb{Z}$ classification
\cite{2017PhRvB..96o5118R}.
With interactions, the topological classification of symmetry class BDI is $\Omega^{{\it Pin}_-}_2(pt)=\mathbb{Z}_8$
\cite{2010PhRvB..81m4509F,2011PhRvB..83g5103F},
which ``misses'' (fails to detect) unitaries with non-zero chiral unitary index.
We however do not dig into this issue further,
as our main focus in this paper is on
unitaries with time-reversal symmetry, for which
the chiral unitary index vanishes.
As we will see, $JR$ symmetry does not seem to play any
role for the case of the time-reversal symmetric model.

\subsection{The two copy theory with time-reversal symmetry}

The shift operator $S$ is odd under time-reversal symmetry.
In order to construct a time-reversal symmetric model of our interest,
we introduce two copies of the 2d chiral topological Floquet model with opposite chiralities.
We use $\uparrow/\downarrow$ to label these two copies.
At the boundary, this time-reversal symmetric model realizes the boundary unitary $U=S_{\uparrow}S^{\dag}_{\downarrow}$,
where $S_{\uparrow/\downarrow}$ is the shift operator that acts exclusively on the first/second copy.
The boundary unitary $U$ acts on the boundary Majorana fermion operators as
\begin{align}
  \label{Majorana 2 copies}
  U \lambda_{\uparrow x} U^{\dag} = \lambda_{\uparrow x+1},
  \quad
  U \lambda_{\downarrow x} U^{\dag} = \lambda_{\downarrow x-1}.
\end{align}
The model is symmetric under the following time-reversal,
\begin{align}
  T \lambda_{\uparrow x} T^{-1} = \lambda_{\downarrow x},
  \quad
  T \lambda_{\downarrow x} T^{-1} = \epsilon \lambda_{\uparrow x},
  \quad
  T^2 = \epsilon^F.
\end{align}
where $(-1)^F=\prod_x (i \lambda_{\uparrow x} \lambda_{\downarrow x})$ is the total fermion number parity operator;
$\epsilon=\pm 1$ distinguishes two cases, symmetry class DIII (BDI) with
$\epsilon= -1 (+1)$. 
Then, the unitary is time-reversal symmetric in the sense that 
$
T U T^{-1}
= T S_{\uparrow} T^{-1} \cdot T S^{\dag}_{\downarrow} T^{-1}
=
\epsilon
S^{\ }_{\downarrow}
S^{\dag}_{\uparrow}
=
\epsilon
(S^{\ }_{\uparrow}S^{\dag}_{\downarrow})^{\dag}
=
\epsilon U^{\dag}.
$
In particular, when $\epsilon=-1$ (DIII), the operator algebra between
$U$ and $T$ is non-trivial, while it is trivial when $\epsilon=+1$ (BDI).


Below, we will try to detect the non-triviality (quantum anomaly)
of the above time-reversal symmetric unitary
when $\epsilon=-1$ (DIII). 
At the non-interacting level, 
Floquet unitaries 
in symmetry class DIII are classified by $\mathbb{Z}_2$
\cite{2017PhRvB..96o5118R}.
When we enforce time-reversal symmetry of class DIII $(\epsilon=-1)$,
by the operator-state map, we consider a short-range entangled state
in the doubled Hilbert space. The relevant symmetry group is
$\mathscr{G}=\mathbb{Z}^f_2 \times \mathbb{Z}_2$
where $\mathbb{Z}^f_2$ represents the fermion number parity conservation, 
and $\mathbb{Z}_2$ is the KMS symmetry ($JT$ symmetry). 
Such short-range entangled states in (1+1)-dimensions are classified by $\Omega^{Spin}_2(B\mathbb{Z}_2) = \mathbb{Z}^2_2$
\cite{Kapustin_2015}.
In the following, we will show explicitly that the above unitary is a
non-trivial element of this group.
Additionally, we also study the operator entanglement spectrum of $U$.
We then find two zero modes which form a doublet under the KMS symmetry.

Let us start by applying the operator-state map 
to the unitary \eqref{Majorana 2 copies}.
We denote the Majorana fermion operators 
acting
on the in and out Hilbert spaces by $\{\lambda_{i/o, \uparrow/\downarrow,x}\}$.
In the doubled Hilbert space, we introduce time-reversal $T$ acting on the fermion operators as
\begin{align}
  T\, \Lambda\, T^{-1}
  =
  \left[
   \mathbbm{1}_2\otimes (i\sigma_2)
        \otimes \mathbbm{1}_L
        \right]\,
        \Lambda,
\end{align}
where
$\Lambda =
\left(
  \lambda_{i\uparrow x},
  \lambda_{i\downarrow x},
  \lambda_{o\uparrow x},
  \lambda_{o\downarrow x}
\right)^T
$
is a $4L$ component vector with $x$ taking values from $1$ to $L$. $\mathbbm{1}_2, i\sigma_2,\mathbbm{1}_L$ act on in/out, spin and position degree of freedoms, respectively.


We choose, as the reference state $|\Omega\rangle\!\rangle$, the ground state of
the quadratic Hamiltonian:
\begin{align}
  \mathbb{H}_0 &= {i} \sum_x
  \left(
  \lambda_{i \uparrow x} \lambda_{o\uparrow x}
 - 
  \lambda_{i \downarrow x} \lambda_{o \downarrow x}
  \right)
  \nonumber \\
  &=
   \frac{i}{2}
    \Lambda^T
      \left(
      \begin{array}{cc}
        0 & (\sigma_3 \otimes \mathbbm{1}_L) \\
        -(\sigma_3 \otimes \mathbbm{1}_L) &0
        \end{array}
      \right)
  \Lambda.
\end{align}
One can check easily $T \mathbb{H}_0 T^{-1}=\mathbb{H}_0$.
We next construct the state $|U\rangle\!\rangle$.
Let $U$ be a generic Gaussian unitary operator
that acts on the fermion operators as
\begin{align}
  U
  \left(
  \begin{array}{c}
    \lambda_{\uparrow } \\
    \lambda_{\downarrow }
    \end{array}
  \right)
  U^{\dag} = \mathcal{Q} 
  \left(
  \begin{array}{c}
    \lambda_{\uparrow } \\
    \lambda_{\downarrow }
    \end{array}
  \right)
\end{align}
where ${\cal Q}$ is 
a $2L\times2L$ real orthogonal matrix, acting on ``in" and ``out" space in the same way.
When $U$ is time-reversal symmetric,   
$\mathcal{Q}$ satisfies 
$
(-i\sigma_2) {\cal Q} (i\sigma_2) = {\cal Q}^T
$.
The state $|U\rangle\!\rangle$ can then be thought of as the ground state of
the parent Hamiltonian
\begin{align}
    \mathbb{H}
    &=
    \frac{i}{2}
    \Lambda^T \, \mathcal{K} \, \Lambda,
      \quad
      \mathcal{K}
      =
       \left(
         \begin{array}{cc}
           0 & (\sigma_3\otimes \mathbbm{1}_L) {\cal Q} \\
           -{\cal Q}^T (\sigma_3\otimes \mathbbm{1}_L) & 0
         \end{array}
       \right).
\end{align}
For the case of our interest, this reduces to 
\begin{align}
  \mathbb{H} = {i} \sum_x
  \left(
  \lambda_{i,\uparrow, x} \lambda_{o,\uparrow, x+1}
 - 
  \lambda_{i,\downarrow, x} \lambda_{o,\downarrow, x-1}
  \right).
\end{align}
The parent Hamiltonian is invariant under
the following unitary operation:
\begin{align}
  &
 (JT)\, \Lambda\, 
  (JT)^{-1}
  =
  \left[
  \left(
  \begin{array}{cc}
    0 & \sigma_1 \\
    \sigma_1 & 0
  \end{array}
  \right)
               \otimes \mathbbm{1}_L
               \right]
               \Lambda
\end{align}
as one can check easily: 
\begin{align}
  &
    \left[
  \left(
  \begin{array}{cc}
    0 & \sigma_1 \\
    \sigma_1 & 0
  \end{array}
  \right)
               \otimes \mathbbm{1}_L
               \right]
               \cdot
               \mathcal{K}
               \cdot
                                       \left[
  \left(
  \begin{array}{cc}
    0 & \sigma_1 \\
    \sigma_1 & 0
  \end{array}
  \right)
               \otimes \mathbbm{1}_L
               \right]
      =
               \mathcal{K},
\end{align}
by using the time-reversal symmetry of $\mathcal{Q}$.
This operation can be understood as
the composition of the modular conjugation
\begin{align}
  &
    J\,
    \Lambda\,
  J^{-1}
  =
  \left[
  \left(
  \begin{array}{cc}
    0 & -\sigma_3 \\
    -\sigma_3 & 0
  \end{array}
  \right)
                \otimes \mathbbm{1}_L
                \right]
                \Lambda,
\end{align}
and time-reversal.
We can check that $J$ leaves $\mathbb{H}_0$ invariant,
$J \mathbb{H}_0 J^{-1}=\mathbb{H}_0$. 
(Also, while $\mathbb{H}$ is not invariant under $T$ nor $J$,
as we checked, $JT$ is a symmetry of $\mathbb{H}$.)
%
Finally, 
an antiunitary operation 
\begin{align}
  &
    (JR)\,
    \Lambda\,
  (JR)^{-1}
  =
  \left[
  \left(
  \begin{array}{cc}
    \mathbbm{1}_2 & 0 \\
    0 & -\mathbbm{1}_2
  \end{array}
  \right)
        \otimes \mathbbm{1}_L
        \right]
        \Lambda,
\end{align}
leaves $\mathbb{H}$ invariant, and acts as chiral symmetry,
\begin{align}
  \left[
  \left(
  \begin{array}{cc}
    \mathbbm{1}_2 & 0 \\
    0 & -\mathbbm{1}_2
  \end{array}
  \right)
        \otimes \mathbbm{1}_L
        \right]
        \cdot
        \mathcal{K}
        \cdot 
        \left[
  \left(
  \begin{array}{cc}
    \mathbbm{1}_2 & 0 \\
    0 & -\mathbbm{1}_2
  \end{array}
  \right)
        \otimes 
        \mathbbm{1}_L
        \right]
        =
        -\mathcal{K}.
\end{align}

\paragraph{Gauging the KMS symmetry}

Let us now gauge the KMS ($JT$) symmetry.
Specifically, we can introduce the background KMS gauge field,
such that we twist the temporal and/or spatial boundary conditions.
As for the temporal twisting, we note, $T U T^{-1} U = (-1)$ and hence
\begin{align}
  \langle \!\langle
  U|
  (JT)
  |U\rangle\! \rangle
  =
  (-1).
\end{align}
This quantity can be interpreted as a partition function on $T^2$
with the twisted temporal direction by the KMS symmetry, and periodic spatial boundary condition.
This confirms that the unitary is a non-trivial element of the $\Omega^{Spin}_2(B\mathbb{Z}_2) = \mathbb{Z}^2_2$ classification.

Similarly, the spatial boundary condition can also be twisted by the KMS symmetry.
To this end, it is convenient to go to the basis that diagonalizes $JT$; we introduce
\begin{align}
  \eta_{\pm} = \lambda_{i\uparrow } \pm \lambda_{o\downarrow },
  \quad
  \xi_{\pm} = \lambda_{i\downarrow } \pm \lambda_{o\uparrow }
\end{align}
The action of $JT$ on these rotated Majorana operators are diagonal:
\begin{align}
  (JT) \eta_{\pm} (JT)^{-1} = \pm \eta_{\pm},
  \quad
  (JT) \xi_{\pm} (JT)^{-1} = \pm \xi_{\pm}.
\end{align}
In terms of these operators, the parent Hamiltonian is written as
\begin{align}
  \mathbb{H} \propto i \sum_x 
  \left(
  \eta_{+, x+1} \xi_{+, x}
  -
  \eta_{-, x+1} \xi_{-, x}
  \right).
\end{align}
Then, twisting spatial boundary condition by $JT$ affects only the minus sector.
In other words, combined with the fermion number parity $(-1)^F$,
we can give different boundary conditions to each sector independently. 
The state $|U\rangle\!\rangle$ can be factorized as $|U \rangle\!\rangle_{pq}
= |U_+\rangle\!\rangle_p |U_- \rangle\!\rangle_q$,
where $p,q$ denote the spatial boundary condition for each sector,
and can be either periodic boundary condition (``Ramond'' boundary condition, $r$),
or anti-periodic boundary condition
(``Neveu-Schwarz'' boundary condition, ${\it ns}$).
The state in the sector twisted by the KMS symmetry is $|U\rangle\!\rangle_{JT}=
|U_+\rangle\!\rangle_{r} |U_-\rangle\!\rangle_{{\it ns}}$.
Then, we see, for example,
\begin{align}
  &
  { }_{JT}\langle \! \langle U| (-1)^F | U \rangle\!\rangle_{JT}
    \nonumber \\
  &\quad =
  { }_{r}\langle \! \langle U_+| (-1)^{F_+} | U_+ \rangle\!\rangle_{r}
    \cdot
  { }_{ns}\langle \! \langle U_-| (-1)^{F_-}| U_- \rangle\!\rangle_{ns}
    \nonumber \\
  &\quad =
    Z[T^2,(r,r)] \cdot Z[T^2,(r,{\it ns})]
    = -1,
\nonumber \\
  &
  { }_{JT(-1)^F}\langle \! \langle U| (-1)^F | U \rangle\!\rangle_{JT(-1)^F}
    \nonumber \\
  &\quad =
  { }_{ns}\langle \! \langle U_+| (-1)^{F_+} | U_+ \rangle\!\rangle_{ns}
    \cdot
  { }_{r}\langle \! \langle U_-| (-1)^{F_-}| U_- \rangle\!\rangle_{r}
    \nonumber \\
  &\quad
    = Z[T^2,(r,{\it ns})] \cdot
    Z[T^2,(r,r)] 
    = -1,
\end{align}
where $Z[T^2, (a,b)]$ is
the torus partition function of (1+1)d topological superconductors
(the Kitaev chain in its non-trivial phase)
in the presence of temporal and spatial boundary conditions $(a,b)$;
$Z[T^2, (a,b)]=-1$ for $(a,b)=(r,r)$,
and 
$Z[T^2, (a,b)]=1$ otherwise
\cite{2017PhRvB..95t5139S}.
We once again confirm that the state $|U\rangle\!\rangle$ is non-trivial in the presence of time-reversal.
The above example represents
the non-trivial element 
$(-1,-1)\in \Omega^{Spin}_d(B\mathbb{Z}_2) = \mathbb{Z}^2_2$
\cite{2015JHEP...12..052K}.


\paragraph{Boundary analysis}

The anomalous properties of $U$ can also be detected by studying 
the boundary excitations or
entanglement spectrum of $|U\rangle\!\rangle$.
Here, we follow \cite{2011PhRvB..83g5103F} to analyze
symmetry actions on the boundary excitations.
When the system (parent Hamiltonian) is cut, excitations
at the boundary are built out of unpaired Majorana fermion operators:
$\lambda_{i,\downarrow}, \lambda_{o,\uparrow}$.
We can then study the algebra of symmetry operators
within the boundary Hilbert space.
For example, 
$JR$, which sends
$\lambda_{i\downarrow} \to +\lambda_{i \downarrow}$
and
$\lambda_{o\uparrow} \to -\lambda_{o \uparrow}$,
can simply be identified as
the complex conjugation, $JR = K$,
if we construct the Fock space by
forming the fermion creation (annihilation) operator
$\lambda_{i\downarrow} \pm i \lambda_{o\uparrow}$
\cite{2011PhRvB..83g5103F}.
$JR$ does not show any anomalous behaviors,
$(JR)^2=1$, as expected since
the unpaired Majorana fermion modes 
$\lambda_{i,\downarrow}$
and
$\lambda_{o,\uparrow}$
carry opposite topological charges.
Proceeding to $JT$ and the fermion number parity,
they can be constructed explicitly as
(see \cite{2015PhRvB..91s5142C, 2017JPhA...50D4002C}
for similar analysis)
\begin{align}
  JT= \frac{e^{i\delta}}
  {\sqrt{2}} (\lambda_{i,\downarrow}+\lambda_{o,\uparrow}),
    \quad
  (-1)^F = (i \lambda_{i,\downarrow}\lambda_{o,\uparrow}),
\end{align}
where the phase 
$e^{i\delta}$ can be chosen such that $(JT)^2=1$,
$e^{2i\delta}=1$.
Now, the commutator between $JT$ and the fermion number parity is
\begin{align}
  (-1)^F (JT) (-1)^F = (-1) (JT).
\end{align}
The projective phase factor $(-1)$ indicates a $\mathbb{Z}_2$ anomaly.
%
%
%

\subsection{Two spatial dimensions}

One can analyze unitary operators of higher-dimensional Majorana fermion systems with time-reversal.
As an example, let us consider $2$ spatial dimensions,
and impose time-reversal symmetry which squares to $(-1)^F$ (class DIII).
Let us once again assume the unitary condition ($JR$ symmetry) does not play any role. 
Then, with the operator-state map, the relevant symmetry group is $\mathbb{Z}^f\times \mathbb{Z}_2$,
where $\mathbb{Z}_2$ is the KMS symmetry ($JT$).
Non-trivial fermionic SPT phases with this symmetry are classified by $\Omega^{Spin}_3(B\mathbb{Z}_2) = \mathbb{Z}_8$
\cite{Ryu_2012, Qi_2013, 2017PhRvB..95t5139S}.
The generating manifold
is $\mathbb{R}P^3$.
The corresponding topological invariant can be constructed
by using partial symmetry transformation acting on a finite subregion of the space
\cite{2017PhRvB..95t5139S}.
Specifically, we can consider the partial KMS symmetry,
combined with $\pi$ spatial rotation $R_{\pi}$,
that acts only on a sub region of the total system,
which we take as a disk $D$.
Therefore, the following expectation value 
\begin{align}
  \langle \! \langle U |
  (JT\cdot R_{\pi})_{D} |U\rangle \! \rangle
  \sim
  e^{ \frac{2\pi \nu i}{8}},
  \quad
  \nu \in \mathbb{Z},
\end{align}
detects the $\mathbb{Z}_8$ classification.
Diagramatically
it can be represented as
$$
\begin{tikzpicture}[thick, scale=.4]
  \begin{scope}
    \draw (0,1) -- (0,0) -- (1,-0.5) -- (2,0) -- (2,3);
    \draw (-1,1) -- (-1,0) -- (1,-1) -- (3,0) -- (3,3);
    \draw (0,2) -- (0,3);
    \draw (-1,2) -- (-1,3);
    \node at (-0.5,1.5) {$U$};
    \draw (-1.5,1) rectangle (0.5,2);
    \draw (0,3) -- (2,4);
    \draw (2,3) -- (0,4);
    \draw (0,4) -- (0,5);
    \draw (-1,3) -- (-1,5);
    \draw (-1.5,6) rectangle (0.5,5);
    \node at (-0.5,5.5) {$U^{\dag}$};
    \draw (-1,6) -- (-1,7) -- (1, 8) -- (3,7) -- (3,3);
    \draw (0,6) -- (0,7) -- (1,7.5) -- (2,7) -- (2,4);
  \end{scope}
\end{tikzpicture}
$$
In terms of the original unitary operator,
this quantity may be obtained by
taking the partial transpose of
the unitary with respect to the disk $D$,
\begin{align}
  \mathrm{Tr}
  \left[
  U^{\dag}
  \cdot
  (R_{\pi})_D U^{T_D} (R^{-1}_{\pi})_D
  \right]
\end{align}
where $A^{T_D}$ represents the partial transpose
of an operator $A$ with respect to $D$.
(Here, we need to use
partial transpose for fermionic systems,
as explained in \cite{2019PhRvA..99b2310S,2017PhRvB..95p5101S,2018PhRvB..98c5151S}.)

We close this section with one remark.
There is an isomorphism (Smith isomorphism)
between $\Omega^{{\it Spin}}_{d+1}(B\mathbb{Z}_2)$ and $\Omega^{{\it Pin}_-}_{d}(pt)$.
This means the classification of boundary unitary in DIII in
$d+1$ spacetime dimension
(= classification of topological Floquet unitary in DIII in $d+2$)
is equivalent to 
the classification of static SPT phases in BDI in $d$
spacetime dimension.
This is consistent since the ``Bott clock'' differs by two
($\cdots$ AI, BDI, D, DIII $\cdots$).

\section{Anomalous unitary operators with $U(1)$ and discrete symmetries}
\label{Anomalous unitary operators with $U(1)$ and discrete symmetries}

\subsection{Generalities}

In this section, we consider topological/anomalous unitary time-evolution operators of charged fermion systems. 
This means that we have particle number conserving symmetry $U(1)$ symmetry, $
e^{ i\theta Q} U e^{ - i\theta Q} =U
$
where $Q$ is the $U(1)$ charge.
In addition, we also discuss various discrete symmetries;
they act on unitaries
as in \eqref{unitary sym} and \eqref{a-unitary sym}.

The $U(1)$ symmetry can be gauged, and we can consider unitary operators
in the presence of background $U(1)$ gauge field, $U(A)$.
In this paper, we focus on time-independent, spatial components of
the $U(1)$ gauge fields,
$A_{i}(\mathbf{r})$ \cite{2019arXiv190803217G}.
To detect topological/anomalous properties of the unitary operator,
we will consider the operator algebra among the unitary symmetries in the
presence of the background gauge field,    
\begin{align}
  \label{UV comm with A}
  V^{\ }_g U(A) V_g^{-1}  = e^{ i \phi_g(A)} U(g\cdot A)^{s(g)},
\end{align}
analogous to 
\eqref{unitary sym} and \eqref{a-unitary sym}.
Here, $s(g)=1$ or $s(g)=-1=\dag$
when $V_g$ is a unitary or anti-unitary symmetry, respectively, and again we allow a possible ``projective'' phase factor. 
$g\cdot A$ represents the background gauge field transformed by symmetry $g$.
For example, $g\cdot A_i(\mathbf{r}) = -A_i(\mathbf{r})$ for particle-hole or time-reversal symmetry.
As before (c.f., Sec.\ \ref{Gauging the KMS symmetry}), 
when an antiunitary symmetry squares to
the identity (possibly up to the fermion number parity $(-1)^F$),
the projective phase $e^{i \phi_g(A)}$ obeys a condition,
$
e^{i \phi_g(g\cdot A)} = e^{ - i \phi_g(A)}e^{i\gamma}
$,
where $\gamma=\pi$ when $V^2_g=(-1)^F$, and $U(A)$ is fermion number
parity odd, and $\gamma=0$ otherwise. 

In addition to the algebraic relation \eqref{UV comm with A},
another closely-related object of our interest is the Schwinger-Keldysh trace:
\begin{align}
  \label{SK trace def}
  Z(A_1, A_2)
  &=
    \mathcal{N}^{-1}
    \mathrm{Tr}\,
    \left[U(A_2)^{\dag}U(A_1) \right]
  \nonumber \\
  &=
  \langle \! \langle U(A_2)|U(A_1) \rangle\! \rangle.
\end{align}
$Z(A_1, A_2)$ is the effective response theory (partition function)
obtained by integrating over matter degrees of freedom.
The Schwinger-Keldysh trace satisfies a couple of constants/conditions, such as 
\begin{itemize}
\item Schwinger-Keldysh symmetry:
\begin{align}
  Z(A, A) =1
\end{align}

\item Reality condition:
\begin{align}
  Z(A_1, A_2)^* = Z(A_2, A_1)
\end{align}
\end{itemize}
These conditions follow directly from \eqref{SK trace def}.
There are also other conditions, in particular, in the presence of symmetries \cite{Glorioso:2016gsa}.
From \eqref{UV comm with A},
\begin{align}
  &
  \mathrm{Tr}
  \big[V^{\ }_g U(A_1) U(A_2)^{\dag} V^{-1}_g \big]
    \nonumber \\
  &\quad
  =
    e^{i [\phi_g(A_1)-  \phi_g(A_2)]}\, 
  \mathrm{Tr}
  \big[
  U(g\cdot A_1)^{s(g)} U(g\cdot A_2)^{-s(g)}
  \big],
\end{align}
we read off
\begin{align}
  &Z(A_1, A_2)
  = e^{i s(g)[\phi_g(A_1)-  \phi_g(A_2)]}\,Z(g\cdot A_1, g\cdot A_2).
\end{align}
The phase factor $e^{i [\phi_g(A_1)- \phi_g(A_2)]}$ is an anomaly in the sense that it 
represents the violation of the naive relation 
$Z(A_1,A_2)= Z(g\cdot A_1, g\cdot A_2)$
expected from the symmetry.

In what follows, we discuss some examples. We consider a series of unitaries in odd spatial dimensions, which, roughly speaking, realize chiral (Weyl) fermions in their single-particle quasi-energy spectrum in momentum space. For example, their single-particle unitaries are given as
$
\mathcal{U}(k_x) = e^{ ik_x}
$ (1d),
$
\mathcal{U}( \mathbf{k}) \sim e^{ i \mathbf{k}\cdot \boldsymbol{\sigma}}
$ (3d),
etc.
These unitaries can be realized as boundary unitaries of
bulk topological Floquet unitaries in one higher dimensions.
The Schwinger-Keldysh trace for these unitaries is given in terms of
topological terms, such as Chern-Simons terms (boundary) and theta terms (bulk).
One of the key questions here is the interplay of these topological terms and
discrete symmetries.

\subsection{Example 1: (1+1)d with $C$}

Let us start with the (1+1)d anomalous unitary,
which is simply a lattice translation operator.
We consider a one-dimensional lattice.
At each site $x$ on the lattice,
we consider complex fermion creation/annihilation operators,
which satisfy the canonical anticommutation relation,
$\{\psi^{\ }_x, \psi^{\dag}_y\}=\delta_{xy}$.
The unitary operator of our interest is the shift operator:
\begin{align}
  \label{def complex shift op}
  U \psi_x U^{-1} = \psi_{x+1}.
\end{align}
The unitary respects the particle number conserving $U(1)$ symmetry,
$e^{i\theta Q}U e^{-i \theta Q}=U$
($\theta \in [0,2\pi]$),
where $Q = \sum_x \psi^{\dag}_x \psi^{\ }_x$ is the total charge.
This unitary arises as a boundary unitary
of a topologically non-trivial 2d Floquet drive
\cite{rudner2013anomalous}.
\footnote{
It can also be viewed as an example of topologically non-trivial non-hermitian Hamiltonian with a point gap.}

As noted in \cite{2019arXiv190803217G},
the unitary operator is invariant under particle-hole symmetry
which is a unitary on-site symmetry defined by
\begin{align}
  \label{shift def cmplx fermion}
  C \psi^{\ }_x C^{-1} =  \psi^{\dag}_x.
\end{align}
(The relevant Altland-Zirnbauer symmetry class is D,
but this case should be distinguished from superconductor realizations of symmetry class D).


In Ref.\ \cite{2019arXiv190803217G},
it was noted that the (bulk and boundary) unitary operators
are symmetric under particle-hole symmetry $C$, $C U C^{-1}\equiv U$,
up to a projective phase for the boundary unitary.
In the presence of the background $U(1)$ gauge field,  
we expect that $C U(A) C^{-1}$ is equivalent to $U(-A)$, $C U(A) C^{-1}\equiv U(-A)$.
While for the bulk without a boundary there is no projective phase
$C U_{{\it bulk}}(A) C^{-1} = U_{{\it bulk}}(-A)$,
one can verify by a direct calculation that
a projective phase exists for the boundary unitary, and it
is given by the one-dimensional Chern-Simons term (Wilson loop),  
\begin{equation}
  \begin{gathered}
    \label{CUC A}
    C\, U(A)\, C^{-1} = e^{ i {\it CS}_1(A)}\, U(-A),
    \\
    {\it CS}_1(A) = \oint dx\, A_x(x)
    =
    \oint A.
  \end{gathered}
\end{equation}
(Possibly up to a phase that is independent of $A$ -- see below.)
We will provide the derivation of the projective phase shortly.
By taking the trace and 
using the operator-state map,
the anomalous relation \eqref{CUC A} 
leads to 
\begin{align}
  e^{ i {\it CS}_{1}(A)}
  &=
  \mathcal{N}^{-1}
  \mathrm{Tr}
  \left[
  C U(A) C^{-1} U^\dag(-A)
  \right]
  \nonumber \\
  &
  =
  \langle \! \langle
  U(A)|\, C \, | U(-A)
  \rangle\!\rangle.
\end{align}
This can be interpreted as the path integral on two-dimensional spacetime with
twisted temporal boundary condition by $C$.

The anomalous algebra \eqref{CUC A} also leads to,
for the ratio of the Schwinger-Keldysh partition functions,
\begin{align}
  \label{SK ratio 1d}
  \frac{Z(-A_1, -A_2)}{Z(A_1, A_2)}
  =
  e^{ -i \oint (A_1- A_2)}
  \neq
  1,
\end{align}
consistent with the result in Ref.\ \cite{2019arXiv190803217G}. 
Furthermore, in Ref.\ \cite{2019arXiv190803217G} it was found that 
the partition function of the corresponding bulk dynamics,
defined on an open spatial manifold with a boundary, 
also picks up a phase under particle-hole symmetry but this has opposite sign:
$Z_{{\it bulk, open}}(-A_1, -A_2)/ Z_{{\it bulk, open}}(A_1, A_2) = e^{ i \oint (A_1-
  A_2)}$.
The total partition function is therefore invariant under $C$. This is the anomaly inflow
%
for the mixed anomaly between the particle-hole and $U(1)$ symmetries. This
consideration extends to the $\mathsf{p},\mathsf{q}$ drives studied
in Ref.\ \cite{2019arXiv190803217G},
in which case (\ref{SK ratio 1d}) becomes $Z(-A_1, -A_2)/ Z(A_1, A_2) = e^{- i
  \theta_{\mathsf{p},\mathsf{q}}/2\pi\oint (A_1- A_2)}$,
where $\theta_{\mathsf{p},\mathsf{q}}$ is an integer multiple of $\pi$ and depends only on $\mathsf{p}/\mathsf{q}$.
Note also that the anomalous relation \eqref{CUC A} leads to 
${\mathrm{Tr}\, U(A)}/{\mathrm{Tr}\, U(-A)}
  =
  e^{i \oint A}$,
which was also verified in Ref.\ \cite{2019arXiv190803217G}.
\footnote{
  We also note that
  the Schwinger-Keldysh trace itself (not the ratio)
  was computed in Ref.\ \cite{2019arXiv190803217G} 
  both for (2+1)d bulk and (1+1)d boundary. In long-wave length limit, the bulk trace is given by
  \begin{align}
    &
      Z_{{\it bulk}}(A_1, A_2)
      \sim
      \exp
      \left[
      \frac{i\theta}{2\pi}
      \int d^2x\, \varepsilon_{ij} \partial_i (A_{1j}-A_{2j})
      \right],
      \label{footeq} 
  \end{align}
  with $\theta=\pi$. There is no such limit for the boundary trace, being the dynamics on the boundary nonlocal.
  Clearly the above expression is consistent with the ratio presented in the main text.
  While (\ref{footeq}) is valid for long wave lengths,
  the ratio presented in the main text for the bulk partition function (as well as (\ref{SK ratio 1d})) is exact.
}

The relation \eqref{CUC A} can be verified 
by a direct calculation for the boundary unitary
of the Floquet topological Anderson insulator,
or
by using the operator-state map and computing 
$\langle \! \langle U(A)|\, C \, | U(-A) \rangle\!\rangle$.

The following results depend on the total number of lattice sites
$(=L)$
being even or odd. 
We note that when we consider 
the 2d topological Floquet system
defined on a cylinder, with two circular boundaries
at its two ends, 
the number of boundary sites per boundary is always even. 
The number of boundary sites can be odd if we consider
different geometries, e.g., a finite 2d square lattice with
a single boundary around it. In the latter case, the boundary
unitary is not a simple shift operator near the corners of the square
lattice.


\subsubsection{Direct calculation}
\label{Direct calculation}

Let us first have a look at the direct calculation.
Following the case of Majorana fermions, we can construct $U$ explicitly: 
$
  U = S_{\lambda} S_{\gamma} (-1)^F
$
where $\psi_x = (\lambda_x+ i \gamma_x)/2$ and $S_{\lambda}$ and $S_{\gamma}$ are
the shift operators for $\lambda_x$ and $\gamma_x$, respectively.
Explicitly,
\begin{equation}
  \begin{gathered}
 U 
   =
  e^{i\delta}
  \lambda_1 \gamma_1
     \Pi_{12}
     \Pi_{23}
     \cdots
     \Pi_{L-1, L} (-1)^F,
 \\ 
    \Pi_{x,x'}
    =
    1 + (\psi^{\dag}_x \psi^{\ }_{x'} - \psi^{\dag}_{x'}\psi^{\ }_x)
    - (n_x - n_{x'})^2. 
    \end{gathered}
\end{equation}
(As before, the phase must be chosen such that $U^L=1$.)
We can then consider to gauge the $U(1)$ symmetry.
This amounts to
$
\psi^{\dag}_x \psi^{\ }_{x'}
 \to
 e^{iA_{x,x'}}\psi^{\dag}_x \psi^{\ }_{x'} 
$.
In addition, we also consider 
\begin{align}
  e^{i (A_{1,2}+A_{2,3}+\cdots + A_{L-1,L})  \psi^{\dag}_1 \psi^{\ }_1 }
  \equiv e^{ i\oint A\, \psi^{\dag}_1 \psi^{\ }_1},
\end{align}
to construct the operator
\begin{align}
  U(A) =
  e^{i\delta}
  e^{ i \oint A\, \psi^{\dag}_1 \psi^{\ }_1}
  \lambda_1 \gamma_1 \Pi_{12}(A)\cdots \Pi_{L-1,L}(A)
  (-1)^F,
\end{align}
as the gauged version of the translation operator.
We can verify
that the gauged version of \eqref{def complex shift op} is given by
\begin{align}\label{uag}
  U(A)\, \psi_x\, U(A)^{-1}
  =
  e^{i A_{x,x+1}}\,
  \psi_{x+1}.
\end{align}
%

The unitary particle-hole transformation $C$,
which acts on the Majorana operators as
$C\lambda C^{-1}=\lambda$ and $C\gamma C^{-1}= - \gamma$,
can also be constructed explicitly: 
\begin{align}
  &
    C = (i \gamma_1\gamma_2)(i\gamma_3 \gamma_4) \cdots (i\gamma_{L-1}\gamma_L).
\end{align}
One can readily check that the following identities hold,
\begin{equation}
  \begin{gathered}
  \label{CUC}
  C U C^{-1} = (-1)^{L+1}U,
    \\
  \mathcal{N}^{-1}
  \mathrm{Tr}\, [C U C^{-1} U^{\dag}] = (-1)^{L+1}.
  \end{gathered}
\end{equation}
The minus sign is indicative of a $\mathbb{Z}_2$ anomaly. 
Now, in the presence of the background gauge field,
\begin{align}
  \label{result CUAC}
  C\, U(A)\, C^{-1} = (-1)^{L+1}\, e^{i \oint A}U(-A).
\end{align}

\subsubsection{Calculation via operator-state map}
\label{Calculation via operator-state map}

Next, let us use the operator-state map,
and calculate
$\langle \! \langle U(A)|\, C \, | U(-A)
\rangle\!\rangle$.
As in the case of Majorana fermion systems
discussed in Sec.\ \ref{Majorana fermion models},
the construction of the reference state proceeds
slightly differently from the bosonic case.
Here, 
the reference state $|\Omega\rangle\!\rangle$
can be conveniently defined as the 
ground state of the parent Hamiltonian
\begin{align}
  \mathbb{H}_0=-\sum_x
  \big(\psi_{i,x}^\dagger \psi_{o,x}+\psi^\dag_{o,x}\psi_{i,x}
  \big),
  \label{eq:ref Hamiltonian}
\end{align}
where
we denote the fermion creation/annihilation operators acting on
the in and out Hilbert spaces
as $\psi^{\dag}_{i,a}/\psi^{\ }_{i,a}$ and $\psi^{\dag}_{o,a}/\psi^{\ }_{o,a}$,
respectively.
Explicitly,
the reference state is given by
\begin{align}
|\Omega\rangle\! \rangle
  =\prod_x\frac{1}{\sqrt{2}}
  \big(\psi^\dag_{i,x}+\psi^\dag_{o,x}\big)
  |0\rangle\!\rangle.
  \label{eq:ref state}
\end{align}
Note that
$|\Omega\rangle\!\rangle$
is given as a superposition of states
of the form
$
|n_{{\it in}}\rangle_i | L-n_{\it in}\rangle_o
=
|n_{{\it in}}\rangle_i (C| n_{\it in}\rangle_o)
$
with $n_{{\it in}}$ being the occupation number
for ``in'' fermions.
Consequently, $|\Omega\rangle\!\rangle$ is invariant under ``vectorial'' $U(1)$ rotations generated by $\exp[ i\theta (Q_i + Q_o)]$,
while it is not under ``axial'' $U(1)$ rotations $\exp[ i\theta (Q_i - Q_o)]$.
Here, $Q_{i/o}=\sum_a \psi^{\dag}_{i/o,a}\psi^{\ }_{i/o,a}$ is the total $U(1)$
charge for the in/out Hilbert space, and $\theta \in [0,2\pi]$.
Alternatively, we could work with a different reference state $|\Omega \rangle\!
\rangle$,
which is invariant under axial $U(1)$ but not under vectorial $U(1)$.
These two choices are simply related by particle-hole transformation $\psi_{o,a}\leftrightarrow \psi^{\dagger}_{o,a}$.
We can introduce a modular conjugation operator $J$ as:
\begin{equation}
  \begin{gathered}
    J \psi_{{\it i},a} J^{-1}= \psi^{\dag}_{{\it o},a},
    \quad
    J \psi_{{\it o},a} J^{-1} = -\psi^{\dag}_{{\it i},a},
    \\
    J|0\rangle\!\rangle = |{\it full} \rangle\!\rangle
    =
    \prod_x
    \big(\psi^\dag_{i,x}\psi^\dag_{o,x}\big)|0\rangle\!\rangle,
    \quad 
    J i J^{-1} = -i.
    \end{gathered}
\end{equation}
One can easily check $|\Omega\rangle\!\rangle$ is invariant under $J$. 

To construct
the mapped state
$|U\rangle\!\rangle$
for the shift operator \eqref{def complex shift op},
we note that $|U\rangle\!\rangle$
is the ground state of
the parent Hamiltonian 
\begin{equation}
  \mathbb{H}=-\sum_x
  \big(\psi_{i,x}^\dagger \psi_{o,x+1}+\psi^\dag_{o,x+1}\psi_{i,x}
    \big).
\end{equation}
Explicitly, $|U\rangle\!\rangle$ is given by
\begin{align}
  |U\rangle\!\rangle
  =\prod_x \frac{1}{\sqrt{2}}
  \big(\psi^\dag_{o,x+1}+\psi^\dag_{i,x}
  \big)|0\rangle\!\rangle.
\end{align}

Particle-hole transformation
\eqref{shift def cmplx fermion}
can be properly extended to act on the doubled Hilbert space, 
\begin{align}
  \begin{gathered}
C\psi^{\ }_{x,i}C^{-1} = \psi^\dagger_{x,i},
  \quad
  C \psi^{\ }_{x,o}C^{-1} = -\psi^\dagger_{x,o},
   \\
  C|0\rangle\!\rangle
  =
|{\it full}\rangle\!\rangle.
\end{gathered}
\end{align}
Note that 
the parent Hamiltonians are invariant under $C$,
$C\mathbb{H}_0 C^{-1}=\mathbb{H}_0$,
and
$C\mathbb{H} C^{-1}=\mathbb{H}$,
and so are their ground states.
One can check explicitly
$C|\Omega\rangle\!\rangle=|\Omega\rangle\!\rangle$
and
$C|U\rangle\!\rangle=(-1)^{L+1}|U\rangle\!\rangle$, 
consistent with \eqref{CUC}.

To study the mixed anomaly, we introduce the background $U(1)$ gauge field via
$U_o^\dag\psi^{\ }_{o,x}U^{\ }_o\rightarrow\psi^{\ }_{o,x+1}e^{iA_{x,x+1}}$,
and consider the parent Hamiltonian
\begin{align}
  \mathbb{H}(A)=
  -\sum_x \left(\psi_{i,x}^\dagger \psi^{\ }_{o,x+1}e^{iA_{x,x+1}}
  +\psi^\dag_{o,x+1}\psi^{\ }_{i,x}e^{-iA_{x,x+1}}
  \right).
\end{align}
The ground state is given by 
\begin{align}
  |U(A)\rangle\!\rangle=\prod_x\frac{1}{\sqrt{2}}
  \left(\psi_{o,x+1}^\dag e^{-iA_{x,x+1}}+\psi^\dag_{i,x}
    \right)|0\rangle\!\rangle.
\end{align}
One can then verify explicitly 
\eqref{result CUAC},
$
    C|U(A)\rangle\!\rangle=(-1)^{L+1}e^{-i\oint A}|U(-A)\rangle\!\rangle.
$

\subsection{Example 2: (3+1)d with ${\it CP}$}
\label{Example 2: (3+1)d with CP}

Let us now discuss the (4+1)d bulk topological Floquet unitary
\cite{2019PhRvL.123f6403H, 2018PhRvL.121s6401S}
and its (3+1)d boundary.
The boundary unitary has a single Weyl point
(or multiple Weyl points with non-vanishing total chiralities)
in its single-particle quasi-energy spectrum.
We can discuss ${\it CP}$ symmetry, which leaves
the boundary unitary invariant,
as seen from 
$
{\it CP}:
{\bf k}\cdot \boldsymbol{\sigma}
\to
  \sigma_2 ( {\bf k}\cdot \boldsymbol{\sigma})^T \sigma_2
  =
  -{\bf k}\cdot \boldsymbol{\sigma}
  $,
where $P$ sends ${\bf r}\to -{\bf r}$ (inversion).
The following discussion using ${\it CP}$
applies also to
${\it CR}$ symmetry, where $R$ sends $x\to -x$ (reflection).

Guided by the 1d case \eqref{CUC A},
we postulate the anomalous operator algebra
relation with the three-dimensional Chern-Simons term
${\it CS}_3(A)$, 
\begin{equation}
  \begin{gathered}
    \label{CP U CP with A}
    ({\it CP})\, U(A)\,
    ({\it CP})^{-1} = e^{i {\it CS}_3 (A)}\, U(\tilde{A}),
    \\
    {\it CS}_3(A) =
    \frac{1}{4\pi} \int d^3x\,
    \varepsilon_{ijk} A_i \partial_j A_k
    =
    \frac{1}{4\pi}\int AdA,
  \end{gathered} 
\end{equation}
where $\tilde{A}={\it CP}\cdot A$ is given by $\tilde{A}_i({\bf r})= A_{i}(-{\bf r})$.
As before, by taking the trace and using the operator-state map,
\begin{align}
  \label{CP U CP with A 2}
  e^{ i {\it CS}_3 (A)}
  &=
  \mathcal{N}^{-1}
    \mathrm{Tr} \left[ ({\it CP}) U(A)
    ({\it CP})^{-1} U(\tilde{A})^{\dag}\right]
    \nonumber \\
  &
  =  
  \langle \! \langle U(A) |({\it CP}) | U(\tilde{A}) \rangle\!\rangle,
\end{align}
where ${\it CP}$ is properly extended so that it acts on the doubled Hilbert space.
In addition, analogously to \eqref{SK ratio 1d},
\eqref{CP U CP with A} leads to
\begin{align}
  \label{SK ratio 3d}
  \frac{Z(A_1, A_2)}{Z(\tilde{A}_1, \tilde{A}_2)}
  =
  e^{
  i [{\it CS}_3(A_1)-{\it CS}_3(A_2)]
  }
  \neq
  1.
\end{align}
Here, we note the Chern-Simons term
${\it CS}_3(A)$
flips its sign under ${\it CP}$,
$\int d^3x\, \varepsilon_{ijk} A_i(-{\bf r}) \partial_j A_k(-{\bf r})
=
-
\int d^3x\, \varepsilon_{ijk} A_i({\bf r}) \partial_j A_k({\bf r}).
$
(This is also the case for $CR$.)
We note that \eqref{CP U CP with A}
is consistent with the Schwinger-Keldysh trace for (4+1)d bulk topological
Floquet unitaries (put on a closed spatial manifold)
and their (3+1) boundary unitaries, 
which are given, in the long-wave length limit, as
\cite{2019arXiv190803217G}
\begin{align}
  \label{4d eff action}
  &
 Z_{{\it bulk}}(A_1, A_2)
  \sim
  \exp
  \left[
  \frac{i\theta}{8\pi^2}
    \int
    (dA_1 dA_1 - dA_2 dA_2)
  \right],
  \nonumber \\
  &
 Z_{{\it bdry}}(A_1, A_2)
  \sim
  \exp
  \left[
    \frac{i}{8\pi}
    \int \, (A_1dA_1- A_2dA_2) 
  \right].
\end{align}
While
$Z_{{\it bulk}}(A_1, A_2)/Z_{{\it bulk}}(\tilde{A}_1, \tilde{A}_2)=1$
for the (4+1) bulk systems,
as inferred from the effective action \eqref{4d eff action},
this naive relation is violated at the boundary, 
\eqref{SK ratio 3d}.

Directly confirming \eqref{CP U CP with A}
along the line of Sec.\ \ref{Direct calculation}
is rather difficult, unfortunately.
Alternatively,
similar to what we did in Sec.\ \ref{Calculation via operator-state map},
we can use the operator-state map 
and compute the overlap
$\langle \! \langle U(A) |({\it CP}) | U(\tilde{A}) \rangle\!\rangle$
in \eqref{CP U CP with A 2}.
In particular, we numerically check that
\begin{align}
    \label{CS3 many-body inv}
\langle \!\langle
    U(\Phi_{xy}=2\pi, \gamma_z )|
  {\it CP}
  | U(\Phi_{xy}=2\pi, \tilde\gamma_z)
  \rangle\!\rangle
\sim e^{i \gamma_z}
\end{align}
holds for a lattice implementation of $|U \rangle\!\rangle$ (See Appendix~\ref{app:CS3} for more details).
 Here, 
$
| U(\Phi_{xy}=2\pi, \gamma_z) \rangle\!\rangle
$
is the mapped state
in the presence of
the unit background magnetic flux
piercing the $xy$ plane
$\Phi_{xy}=\oint F_{xy}=2\pi$
and 
the Wilson loop $\gamma_z = \oint A_z$ along $z$-direction.
Note that $\tilde\gamma_z=-\gamma_z$ as a result of $CP$ transformation.
This background gauge field configuration gives rise to ${\it CS}_3(A)=\gamma_z$.
In the limit $\gamma_z=\pi$, the quantity \eqref{CS3 many-body inv}
is essentially 
the same as the $\mathbb{Z}_2$ many-body topological invariant 
for fermionic short-range entangled states
protected by ${\it CP}$ (or ${\it CR}$) symmetry
in (3+1) dimensions
(topological insulators in symmetry class A + ${\it CR}$ with
$({\it CR})^2=1$), introduced in \cite{2018PhRvB..98c5151S}.



\subsection{Comments}

Let us close this section with some comments.

--
First of all,
while we focused here on the anomalous unitaries
preserving $U(1)$ in one and three spatial dimensions
(with $C$ and $CP$ symmetries, respectively),  
we expect that the pattern continues to all higher odd spatial dimensions.
The anomalous operator algebras in higher dimensions
signifying a mixed anomaly between
$U(1)$ and a discrete symmetry
involve higher-dimensional Chern-Simons terms,
$\int A dA \cdots dA$. 
This is analogous to the ``primary series''
of topological insulators/superconductors
in even (odd) spatial dimensions
that are classified/characterized
by an integral topological invariant
and the response Chern-Simons terms ($\theta$ terms) 
\cite{2008PhRvB..78s5424Q,2010NJPh...12f5010R}.
For a given spatial dimension,
they belong to one of the ten Altland-Zirnbauer symmetry classes.

--
There are also topological states that are
outside of the primary series,
and are classified by $\mathbb{Z}_2$ topological invariants
-- they are obtained from the topological states in
the primary series by dimensional reduction
(called ``(first/second) descendants'' in \cite{2010NJPh...12f5010R}).
For anomalous unitaries, we also expect that
there are similar ``descendants''.
For example, 
let us consider unitaries in two spatial dimensions
respecting $U(1)$ and ${\it CR}$ symmetries.
Following \cite{2018PhRvB..98c5151S},
we can construct the $\mathbb{Z}_2$ topological invariant as
\begin{align}
  &
    \frac{
    \langle\!\langle  U(\Phi_{xy}=2\pi)|
    {\it CR}
    | U(\Phi_{xy}=2\pi)
    \rangle\!\rangle
    }
    {
    \langle \! \langle U(\Phi_{xy}=0)|
    {\it CR}
    | U(\Phi_{xy}=0)
    \rangle\! \rangle 
    }
    =
    \pm 1.
\end{align}
Here, the background gauge field configuration is invariant under
${\it CR}$.
An anomalous unitary for which this topological invariant
is non-trivial
should have an even number of Weyl points in its quasi-energy spectrum.

--
There is a close connection between
the effective Schwinger-Keldysh functional
$Z(A_1, A_2) =\langle \!\langle U(A_1)| U(A_2) \rangle\! \rangle$
and the Berry phase
$\oint \langle \!\langle U(A)| U(A+dA) \rangle\! \rangle$.
The relations like \eqref{4d eff action}
can be guessed from
(or at least consistent with)
the Berry phase of the short-range entangled state
$|U(A)\rangle\!\rangle$.
For a short-range entangled state
$|U(A)\rangle\!\rangle$ in the presence of
a spatial background gauge field $A$,
it is known that
that the Berry phase 
is related to the response effective action
\cite{2018PhRvB..98c5151S}. For example, 
for
$\mathcal{U}({\bf k}) \sim e^{ i {\bf k}\cdot \boldsymbol{\sigma}}$, 
the Berry phase is related to the $\theta$ term
in the effective response action, 
\begin{align}
  \label{Berry phase}
  \oint dA\,
  i \langle \! \langle U(A)|\frac{d}{dA}|U(A) \rangle\! \rangle
    &=
      \frac{i \theta}{8\pi^2}
      \int dt d^3x\,
      \varepsilon^{\mu\nu\kappa\lambda}
      \partial_{\mu}A_{\nu}\partial_{\kappa}A_{\lambda}
\end{align}
with $\theta=\pi$.
specializing to the configuration
for which $\varepsilon_{ij}\partial_{i}A_j$ ($i,j=x,y$)
is time-independent,
but changing $A_z$ adiabatically in time,
and further 
discretizing the (adiabatic) time,
$\partial_t A_{\mu} \to A_{1\mu} - A_{2\mu}$,
\eqref{Berry phase} suggests
\begin{align}
  \langle \! \langle U(A_1)|U(A_2) \rangle\! \rangle
 \sim 
  1 + \frac{i\theta}{4\pi^2}
  \int d^3x\, (A_{1z}-A_{2z}) \varepsilon_{ij} \partial_{i} A_{j}
\end{align}
where $A_y =A_{1y}=A_{2y}= (A_{1y}+A_{2y})/2$.
This is consistent with \eqref{4d eff action}.
To summarize, 
we can use the operator-state map and the Berry phase
to ``guess'' the Schwinger-Keldysh response effective action
$Z(A_1, A_2)$
when $A_1$ and $A_2$ are close enough.

\section{Conclusion}

In this paper, we discuss the characterizations
of anomalous unitary time-evolution operators,
that may be
realized on the boundary of bulk topological Floquet systems.
Much the same way as the boundaries of static topological phases
that can be characterized, detected, and classified by
quantum anomalies, we identified quantum anomalies for boundary unitaries.

We close by listing a few open questions and
interesting directions to explore.

--
First, while we focused on quantum anomalies on boundary unitaries,
it is interesting to ask if there is a corresponding
bulk topological field theory.
This problem was explored already in
\cite{2019arXiv190803217G} for the case of background $U(1)$ gauge field. 
For the case of time-reversal symmetric boundary unitaries,
it is interesting to ask if one can write down
a topological field theory for the KMS gauge field.

--
As mentioned in Sec.\ \ref{Gauging unitary symmetries},
anomalous boundary unitaries are characterized
by their algebraic relations with symmetry generators. 
In the presence of $U(1)$ symmetry, we considered gauged versions of
the anomalous operator algebra in
Sec.\ \ref{Anomalous unitary operators with $U(1)$ and discrete symmetries}.
Instead of gauging $U(1)$ symmetry,
it would be interesting to consider
the Lieb-Schultz-Mattis type twist operator,
which has been useful in
various Lieb-Schultz-Mattis type theorems
and can be understood in terms of quantum anomalies.

--
It is interesting to
apply/extend the framework developed in this paper
to other symmetries. 
For example, in
Sec.\ \ref{Anomalous unitary operators with $U(1)$ and discrete symmetries},
we discussed the mixed anomaly between $U(1)$ and discrete symmetric, $C$ and
${\it CP}$.
It would be interesting to discuss mixed anomalies between $U(1)$
and other discrete symmetries.

--
It is also interesting to study ``exotic'' symmetries, such as
$O U O^{-1} = U^{\dag}$ where $O$ is a unitary operator.
The symmetry is called ``many-body spectral reflection symmetry''
or ``unitary time-reflection symmetry''
\cite{2017JSMTE..08.3105C, 2018PhRvL.120u0603I,2018PhRvB..98c5139S}.
By combining with time-reversal, we can also consider 
$O' U O'^{-1}= U$
where $O'$ is an antiunitary operator.
In the doubled Hilbert space, 
when $U=O U^{\dag}O^{-1}$,
the composition of the modular conjugation $J$ with $O$
is an antiunitary symmetry, 
$O J|U\rangle\!\rangle = O|U^{\dag}\rangle\!\rangle
  \equiv |U \rangle\!\rangle$.
$JO$ can be gauged, 
by putting the system on an unoriented spacetime,
$\mathbb{R}P^2$.
It would be interesting to see if
the associated topological invariant
(the partition function on $\mathbb{R}P^2$)
is related to the corresponding
matrix-product-operator
index discussed in \cite{2017JSMTE..08.3105C}.

--
Finally, while our focus in this paper is on unitaries with symmetries,
it is interesting to see if one can understand
the chiral unitary index of 1d unitaries, in terms of quantum anomalies.
A natural candidate is a gravitational anomaly.


%
%
%
%
%
%
%

\acknowledgements

The authors would like to acknowledge insightful discussions
with Yoshimasa Hidaka, Masaru Hongo, and Ryohei Kobayashi.
SR is supported by a grant from the Simons Foundation (Award Number: 566116).

\appendix

\section{Numerical verification of manybody topological invariant in Eq.~(\ref{CS3 many-body inv})}
\label{app:CS3}

In this appendix, we consider a lattice Hamiltonian in the doubled Hilbert space and  numerically show that
the relation (\ref{CS3 many-body inv}) holds.

Recall that the shift operator as a boundary unitary of $(2+1)$d Floquet topological is given by Eq.~(\ref{def complex shift op}). The corresponding transformation in momentum space is then
$
    U \psi_k U^{-1} = e^{ik} \psi_k,
$
where $\psi_k = \sum_{x} \psi_x e^{-ikx}$. Thus, the $(3+1)$d generalization of this boundary unitary becomes 
\begin{align}
 U \psi_{\bf k} U^{-1}= e^{i {\bf k}\cdot \boldsymbol{\sigma}}
 \psi_{\bf k},
 \label{eq: 3d unitary cp}
\end{align}
where $\psi_{\bf k}$ is a two-component fermionic field. Our system is furnished with a $CR$ symmetry which acts as
\begin{align}
(CR) \psi({\bf r}) (CR)^{-1}= \psi^\dag(R {\bf r}),
\end{align}
where $R {\bf r} = (x,-y,z)$ involves a reflection with respect to $xz$ plane.
The corresponding transformation in momentum space reads as
\begin{align}
(CR) \psi_{\bf k} (CR)^{-1}= \psi^\dag_{-R{\bf k}}.
\end{align}
where $-R {\bf k} = (-k_x,k_y,-k_z)$. It is easy to check that the unitary (\ref{eq: 3d unitary cp}) is invariant under $CR$.

To construct the mapped state $|U\rangle\!\rangle$ we use the reference state  (\ref{eq:ref state}) which is the ground state of the Hamiltonian (\ref{eq:ref Hamiltonian}). 
Hence, the state $|U\rangle\!\rangle$ can be obtained as the ground state of the following Hamiltonian
\begin{align}
\label{CS3 continuum H}
\mathbb{H} = -\int d^3k\ 
\Psi_{\bf k}^\dag
\begin{pmatrix}
0 & e^{-i {\bf k}\cdot \boldsymbol{\sigma}} \\
e^{i {\bf k}\cdot \boldsymbol{\sigma}} & 0 
\end{pmatrix}
\Psi_{\bf k},
\end{align}
where $\Psi_{\bf k}^\dag = (\psi_{o,{\bf k}}^\dag,\psi_{i,{\bf k}}^\dag)$. The
proper $CR$ transformations in the doubled Hilbert space are given by
\begin{align}
(CR) \psi^{\ }_{i,{\bf k}} (CR)^{-1} &= \psi^\dagger_{i,-R{\bf k}},
  \quad
   \nonumber \\
  (CR) \psi^{\ }_{o,{\bf k}} (CR)^{-1} &= -\psi^\dagger_{o,-R{\bf k}},
   \nonumber \\
  (CR)|0\rangle\!\rangle
  &=
|{\it full}\rangle\!\rangle.
\label{eq: CR inout}
\end{align}

In order to calculate the quantity (\ref{CS3 many-body inv}), we need to find the ground state in the presence of magnetic field in $xy$-plane and twisted boundary condition in $z$ direction. This requires a real-space implementation of the Hamiltonian. It is more convenient to consider the following Hamiltonian on a cubic lattice
\begin{align} \label{CS3 lattice H}
  \mathbb{H}_{\text{latt}}=& \frac{1}{2} \sum_{\substack{\textbf{r}\\ s=1,2,3}}
  {\Big[} \Psi^\dagger ({\textbf{r}+\hat{x}_s}) (i t \alpha_s - r\beta) \Psi (\textbf{r}) +\text{H.c.} {\Big]} \nonumber \\
&+ m \sum_{\textbf{r}} \Psi^\dagger(\textbf{r})  \beta \Psi (\textbf{r}),
\end{align}
which shares the same low-energy Hamiltonian as that of Eq.~(\ref{CS3 continuum H}).
Here, the Dirac matrices are given by
\begin{align}
\alpha_s &= \tau_y\otimes \sigma_s=\left(\begin{array}{cc}
0 & -i\sigma_s \\ i\sigma_s & 0
\end{array} \right), \\
\beta &= \tau_x\otimes I=\left(\begin{array}{cc}
0 & I \\  I  & 0
\end{array} \right),
\end{align}
where $\tau$ acts on in/out degrees of freedom and $\sigma$ acts on the inner degrees of freedom.

For simplicity, we set $t=r=1$.
Furthermore, the lattice Hamiltonian is also invariant under
${\it CR}$ symmetry defined in Eq.~(\ref{eq: CR inout}). 
We should note that the ground state of the lattice Hamiltonian with a mass term in the range $1<|m|<3$ is topologically equivalent to the mapped unitary $|U\rangle\!\rangle$.

To compute the ground state
$| {\it GS}(\Phi_{xy}=2 \pi, \gamma_z)\rangle$, we modify the hopping terms in the lattice Hamiltonian (\ref{CS3 lattice H}), which we call $\mathbb{H}_{\text{latt}}(
\Phi_{xy}=2 \pi, \gamma_z)$, as follows:
A simple way to prepare a $2\pi$ magnetic flux with uniform magnetic field is to set 
\begin{align} 
A_x(x,y) &= -\frac{2 \pi y}{L_xL_y},\nonumber \\
A_y(x,y) &= \left\{\begin{array}{ll}
0 & (y=1, \dots, L_y-1) \\
\frac{2 \pi x}{L_x} & (y=L_y) \\
\end{array}\right. , 
\label{eq:uniform_magnetic_flux}
\end{align}
where $L_x,L_y$ are the number of sites. 
This gauge configuration leads to a uniform magnetic flux $F(x,y) = A_x(x,y)+A_y(x+1,y)-A_x(x,y+1)-A_y(x,y)=2 \pi/(L_x L_y)$ inserted per unit cell. 
The twisted boundary condition in $z$ direction is implemented as usual via multiplying the hopping amplitudes by a phase factor.
It is important to note that the quantity (\ref{CS3 many-body inv}) is
well-defined since under $CR$ symmetry we have 
\begin{align}
    (CR)\,\mathbb{H}_{\text{latt}}\, (
\Phi_{xy}, \gamma_z) (CR)^{-1}
=\mathbb{H}_{\text{latt}} (
\Phi_{xy}, -\gamma_z).
\end{align}

\begin{figure}[H]
\includegraphics[scale=1.35]{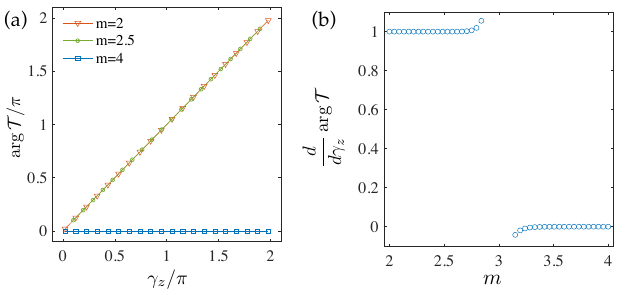}
\caption{\label{fig:CR} Manybody topological invariant for $CR$-symmetric $(3+1)$d boundary unitary (Sec.~\ref{Example 2: (3+1)d with CP}). The invariant was computed via (\ref{invariant to plot})
for the ground state of the lattice Hamiltonian (\ref{CS3 lattice H}) in the doubled Hilbert space. 
The topological phase of the lattice Hamiltonian (i.e., when $2<m<3$) is topologically equivalent to $|U\rangle\!\rangle$. }
\end{figure}

Figure~\ref{fig:CR}(a) shows how the argument of the following quantity
\begin{align}
\label{invariant to plot}
    {\cal T}(\gamma_z)=  \langle
    {\it GS}(\Phi_{xy}=2\pi, \gamma_z )|
  {\it CR}
  | {\it GS}(\Phi_{xy}=2\pi, -\gamma_z)
  \rangle,
\end{align}
 varies as a function of $\gamma_z$. We plotted two values for the mass term in the topological phase and one in the trivial phase. It is evident that in the former case $\arg {\cal T}=\gamma_z$ while in the latter $\arg {\cal T}=0$.  We further check that the linear behavior $\arg{\cal T}=\gamma_z$ is valid within the topological phase (away from the transition point $m=3$ where finite-size effects dominate) in Fig.~\ref{fig:CR}(b). 

Given the topological equivalence between the topological phase of $\mathbb{H}_{\text{latt}}$ and the ground state of $\mathbb{H}$,
we deduce Eq.~(\ref{CS3 many-body inv}).

\bibliographystyle{ieeetr}
\bibliography{gauginganomalousunitary}

\end{document}